\documentclass[nologo,11pt,a4paper]{ETHpaper}

\usepackage{verbatim}  
\usepackage{graphicx}                  
\usepackage[compress]{natbib}     

\usepackage{amsfonts}
\usepackage{array}
\usepackage{amsthm}
\usepackage{amsmath}
\usepackage{caption}
\graphicspath{{fig/}}
\hypersetup{
    colorlinks=false,
    pdfborder={0 0 0},
}

\usepackage{color}
\usepackage{soul}

\newcommand{\mean}[1]{\left\langle #1 \right\rangle} 
\newcommand{\abs}[1]{\left| #1 \right|} 

\renewcommand{\epsilon}{\varepsilon} 
 
\renewcommand*{\=}{{\kern0.1em=\kern0.1em}}
\renewcommand*{\-}{{\kern0.1em-\kern0.1em}} 
\newcommand*{\+}{{\kern0.1em+\kern0.1em}}

\begin{document}
 
\title{Quantifying knowledge exchange in R\&D networks: A data-driven model}
\titlealternative{Quantifying knowledge exchange in R\&D networks: A data-driven model\\
\textit{Journal of Evolutionary Economics} (revised and resubmitted, April 2017)}

\author{Giacomo Vaccario$^{1}$, Mario V. Tomasello$^{1}$, \\ Claudio J. Tessone$^{1,2}$ and Frank Schweitzer$^{1,*}$}

\address{
\bigskip \small
$^1$ETH Zurich, Chair of Systems Design, Department of Management, Technology and \\ Economics, Weinbergstrasse 56/58, CH-8092 Zurich, Switzerland\\
$^2$Universit\"at Z\"urich, URPP Social Networks, Department of Business Administration, \\ Andreasstrasse 15, CH-8050 Zurich, Switzerland\\
\bigskip
$^\ast$Corresponding author; E-mail: fschweitzer@ethz.ch 
}
\authoralternative{G. Vaccario, M. V. Tomasello, C. J. Tessone, F. Schweitzer}

\maketitle

\begin{abstract}
\noindent
We propose a model that reflects two important processes in R\&D activities of firms, the formation of R\&D alliances and the exchange of knowledge as a result of these collaborations. 
In a data-driven approach, we analyze two large-scale data sets extracting unique information about 7500 R\&D alliances and 5200 patent portfolios of firms. 
This data is used to  calibrate the model parameters for network formation and knowledge exchange. 
We obtain probabilities for incumbent and newcomer firms to link to other incumbents or newcomers which are able to reproduce the topology of the empirical R\&D network. 
The position of firms in a knowledge space is obtained from their patents using two different classification schemes, IPC in 8 dimensions and ISI-OST-INPI in 35 dimensions. 
Our dynamics of knowledge exchange assumes that collaborating firms approach each other in knowledge space at a rate $\mu$ for an alliance duration $\tau$. 
Both parameters are obtained in two different ways, by comparing knowledge distances from simulations and empirics and by analyzing the collaboration efficiency $\mathcal{\hat{C}}_{n}$. 
This is a new measure, that takes also in account the effort of firms to maintain concurrent alliances, and is evaluated via extensive computer simulations. 
We find that R\&D alliances have a duration of around two years and that the subsequent knowledge exchange occurs at a very low rate. 
Hence,  
a firm's position in the knowledge space is rather a determinant than a consequence of its R\&D alliances.
{From our data-driven approach we also find model configurations that can be both realistic and optimized with respect to the collaboration efficiency $\mathcal{\hat{C}}_{n}$.
Effective policies, as suggested by our model, would incentivize shorter R\&D alliances and higher knowledge exchange rates.}
\end{abstract}


\section{Introduction}
\label{sec:intro}

The last three decades have been characterized by a growing number of inter-firm alliances, aimed at Research and Development (R\&D) purposes. Albeit this phenomenon has especially affected highly technological industries such as IT, Pharmaceuticals or Medical Supplies  \citep{ahuja2000collaboration,hagedoorn2002inter}, all industrial sectors have simultaneously experienced an increased number of such alliances  \citep{tomasello2016riseandfall}.
Consequently, scholars have investigated the mechanisms behind the formation of R\&D alliances \citep{powell2005network}, the complex networks they generate  \citep{rosenkopf2007comparing,tomasello2014therole}, and the way their evolution can be modeled  \citep{konig2012efficiency,garas2017selection}. 

From a theoretical point of view, it has been shown that firms engage in alliances for several reasons. They can gain access to more and diverse assets  \citep{liebeskind96:_knowl_strat_theor_firm,das2000resource}. Next, alliances foster the exchange of knowledge between firms: by joining their technological resources, firms can actually enlarge their knowledge bases faster than they could do individually  \citep{baum2000don,mowery1998technological,rosenkopf2003overcoming}. 
Finally, firms can share the costs and risks of a project, especially when this is expensive or with uncertain outcome  \citep{hagedoorn00:_resear}. 
All of these aspects result in a learning process of the involved firms, making R\&D alliances an important part of every firm's knowledge management strategy.

The focus of the present study is indeed such a learning process, which we model as a mutual exchange of knowledge occurring after the establishment of an alliance between two firms.
In particular, we develop an agent-based model to investigate the determinants leading to the formation of inter-firm R\&D collaborations and the subsequent emergence of an R\&D network. At the same time, we study the effect that such collaborations have on the technological positions of the involved firms, and we estimate the \textit{performance} of such networked systems, in terms of explored technological space.

The approach that we adopt in our study can be defined as \textit{data-driven modeling}. 
Starting from the empirical evidence, we design a set of realistic and theoretically grounded microscopic interaction rules, which we incorporate in an agent-based model; next, we implement the model through computer simulations, followed by calibration and validation against empirical data.
The fine-tuning of the model parameters gives us not only a deep understanding of the system under examination, but also an indication on how to optimize it. 
The model that we develop here is based on previous empirical findings \citep{tomasello2016riseandfall, hanaki2010dynamics, rosenkopf2007comparing}, and combines two existing agent-based models \citep{tomasello2014therole, tomasello2016knowledge_exchange}, in order to reproduce both the alliance formation and the knowledge exchange process in an R\&D network.

\subsection{Theoretical foundations: knowledge exchange in inter-firm R\&D networks}
\label{sec:intro_knowledge_exchange}

Our agent-based model follows a number of extant works on bounded confidence and continuous opinion dynamics  \citep{axelrod1997dissemination,deffuant2000mixing,degroot1974reaching,hegselmann2002opinion,groeber2009groups}, in particular applied to innovation networks  \citep{fischer2001knowledge,baum2010network}.
In the wake of this previous work, and similar to the model proposed by \citet{tomasello2016knowledge_exchange}, we assume that the collaborating agents are characterized by an evolving knowledge basis, that is affected by the set of alliances in which are involved. However, differently from the studies that have been done so far, our model does not focus on the formation of consensus clusters -- see \citet{axelrod1997dissemination,schweitzer2009nonlinear} in the case of social systems, or \citet{fagiolo2003exploitation} for technology islands, but on the exploration of a {\textit knowledge space} (defined below). 
In addition, our work does not consider the network of R\&D alliances as fixed, but it assumes a dynamically evolving R\&D network, whose topology corresponds to those of empirically observed networks \citep[see][]{tomasello2016riseandfall, gulati2012:_rise_and_fall_of_small_world}.

The knowledge-based view of the firm \citep{fischer2001knowledge} assumes that every company is endowed with a knowledge basis that uniquely identifies its resources and capabilities. In other words, a firm can always be associated with a vector consisting of several components \citep{sampson2007diversity}, each of which represents its level of knowledge in a given area. These vectors can in turn be associated with a metric \textit{knowledge space} in which the collaborations occur. 
Thus, every firm occupies a point in this multi-dimensional space, whose coordinates are given by its knowledge vector.
Such an approach is similar to a more general model  \citep{axelrod1997dissemination}, proposed in the broader context of social influence.
The concept of a metric knowledge space has already been used in one  \citep{groeber2009groups}, and two dimensions  \citep{fagiolo2003exploitation,baum2010network}; here, we generalize this approach to metric spaces of arbitrary dimensionality.

On the other hand, R\&D alliances have been conceptualized by several studies \citep{mowery1998technological, owen2004knowledge, grant2004knowledge, gomes2006alliances} as a means to exchange technological knowledge among firms, and such an idea is at the heart of several agent-based models  \citep{pyka2007,gilbert04:_agent_based_social_simulation_complexity,cowan2007bilateral}. In these models, the agents' knowledge bases become more similar over time, as a consequence of R\&D collaborations.
The speed at which the agents approach each other in the knowledge space represents one of the fundamental parameters of this family of models, and our work is no exception. 
Besides, we rely on the assumption that knowledge spillovers occurring in a R\&D alliance cause the partners to exchange knowledge along every dimension of their knowledge bases, not limiting the transfer to a specific R\&D project that they have in common \citep{baum2010network}.
In other words, we study a scenario in which the two partners approach with respect to all dimensions of the knowledge space.

Finally, we aim at studying the \textit{performance} of the whole collaboration network as a function of the relevant model parameters. To quantify it, we propose a measure that takes into account the global knowledge exploration of the systems. 
I.e., it takes into account the distances in knowledge space traveled by all agents during the evolution of our simulated R\&D network.
In our model, we consider that the knowledge exploration itself is represented by the motion in the knowledge space, which is fully captured by such a measure.
The underlying assumption is that a throughout exploration of the knowledge space is beneficial for the R\&D network, in that it allows the agents to come in contact with many technological opportunities, potentially leading to more frequent innovations \citep{fagiolo2003exploitation}.
Precisely, we make use of an existing performance indicator \citep{tomasello2016knowledge_exchange} and refine it by taking into account the actual number of active collaborations in the system, in order to obtain a more reliable measure.

\subsection{Theoretical foundations: formation of inter-firm R\&D networks}
\label{sec:intro_network}

The extant literature on R\&D networks has shown that two crucial types of mechanisms drive the formation of new R\&D alliances \citep{rosenkopf2008investigating}: \textit{endogenous} mechanisms and \textit{exogenous} mechanisms.
The endogenous mechanisms depend on firms' social capitals which describe the firms' positions in the network, while the exogenous mechanisms are affected by firms' technological and commercial capitals.
Here, we refer to an alliance as ``endogenous'' if it involves a partner that belongs already to the R\&D network. 
While if it involves a partner that does not belong to the R\&D network, we refer to the alliance as ``exogenous''.

Typically, empirical and theoretical studies have focused on the mechanisms driving endogenous and exogenous alliances separately, also called ``network endogeneity'' \citep{KogutWalkerShan1997, powell1996interorganizational, gulati1999interorganizational, garas2017selection} and ``exogenous partner selection'' \citep{BurtStrucHoles92, rosenkopf2001beyond, cowan04:_knowl_dynam_networ_indus}.
However, to explain the observed empirical R\&D network both types of mechanisms are needed. 
As matter of fact, network endogeneity by itself would produce more and more centralized network over time, which does not occur in the real R\&D network \citep{tomasello2016riseandfall}.  
On the other hand, a purely exogenous partner selection would lead to regular network topologies, which also does not occur (a prominent example is represented by the ``monogamous'' networks analyzed by \citet{tomasello2016knowledge_exchange}).
A notable exception is the agent-based model developed by \citet{tomasello2014therole}, which incorporates both endogenous and exogenous rules of alliance formation and successfully reproduce the structure of a real R\&D network.
In fact, the model permits to tune the weight of both endogenous and exogenous mechanisms for alliance formation, and to test the outcome against real data.

Inspired by these works, the agent-based model that we develop in the present study includes all the microscopic rules introduced in \citet{tomasello2014therole}, and combines them with the knowledge exchange rules briefly discussed above. 
Our model allows us to modulate the weight of both endogenous and exogenous mechanisms for alliance formation, and to study the knowledge exchange in R\&D networks.

\subsection{Our contribution}
As mentioned, {we combine}, and extend, two existing agent-based models in a straightforward, yet effective, manner. The model introduced by \citet{tomasello2016knowledge_exchange} represents a first attempt to investigate the process of knowledge exchange occurring in a dynamic collaboration network; it has identified a mechanism of volatile alliances to help the collaborating agents better explore a knowledge space, using the approximation of monogamous (i.e. sparse) collaboration networks.
On the other hand, the model developed by \citet{tomasello2014therole} can realistically reproduce the complex topology of real R\&D networks, but without considering the effect of alliances on the firms' knowledge positions.

The agent-based model we introduce here constitutes an important step toward a general framework that combines two dynamic processes, the formation of alliances and the knowledge exchange in collaboration networks.
The microscopic interaction rules of our model and its calibration involve a two-step procedure that can be described as follows.
The firms form R\&D collaborations based on their network features and their social capital; the model parameters related to these mechanisms are estimated through a comprehensive inter-firm alliance data set. Next, we assume that the formation of each network link causes a process of knowledge exchange between the involved firms, which consequently approach in the knowledge space; the model parameters related to this mechanism are estimated through a second data set on firm patents.
Remarkably, the underlying knowledge space that we consider in our study is defined by real patent classes, allowing for a precise quantification of every firm's technological position. 
In this paper, we also investigate how the dimensionality of the knowledge space impacts our results. 

Our findings point out a predominance of the endogenous network mechanisms (over the exogenous ones) for the alliance formation; in other words, previous network structures and alliance history matter when selecting new collaboration partners. Next, we find that real R\&D alliances have a duration of around two years, and that the subsequent knowledge exchange between the partners occurs at a very low rate. Most of the alliances, indeed, have no consequence on the partners' knowledge position: this suggests that a firm's position -- evaluated through its patents -- is rather a determinant than a consequence of its R\&D alliances. Finally, we investigate the performance of such a network in terms of explored knowledge trajectories, and we check whether the real R\&D network under examination maximizes our proposed performance indicator. 
Interestingly, we find that this is the case: effective policies to obtain an optimized collaboration network -- as suggested by our model -- would incentivize shorter R\&D alliances and higher knowledge exchange rates.

The rest of the paper is organized as follows. Section \ref{sec:data_methods} presents the data sets and the methodology used to build the network, as well as to measure the firms' knowledge positions.
Section \ref{sec:model} describes all the microscopic interaction rules defining our agent-based model.
Sections \ref{sec:network_formation_validation} and \ref{sec:knowledge_exchange_validation} present the results of our computer simulations and the model calibration on the alliance and the patent data sets, respectively. 
In Section \ref{sec:performance}, we introduce a quantification of the collaboration
efficiency and study the optimality of the real R\&D network under examination.
Finally, Section \ref{sec:conclusions} concludes.


\section{Data and Methodology}
\label{sec:data_methods}

\subsection{Network reconstruction, activities and patents}
\label{sec:netw-reconstr-activ}

\begin{figure}[htbp]
  \centering
  \includegraphics[width=0.495\textwidth]{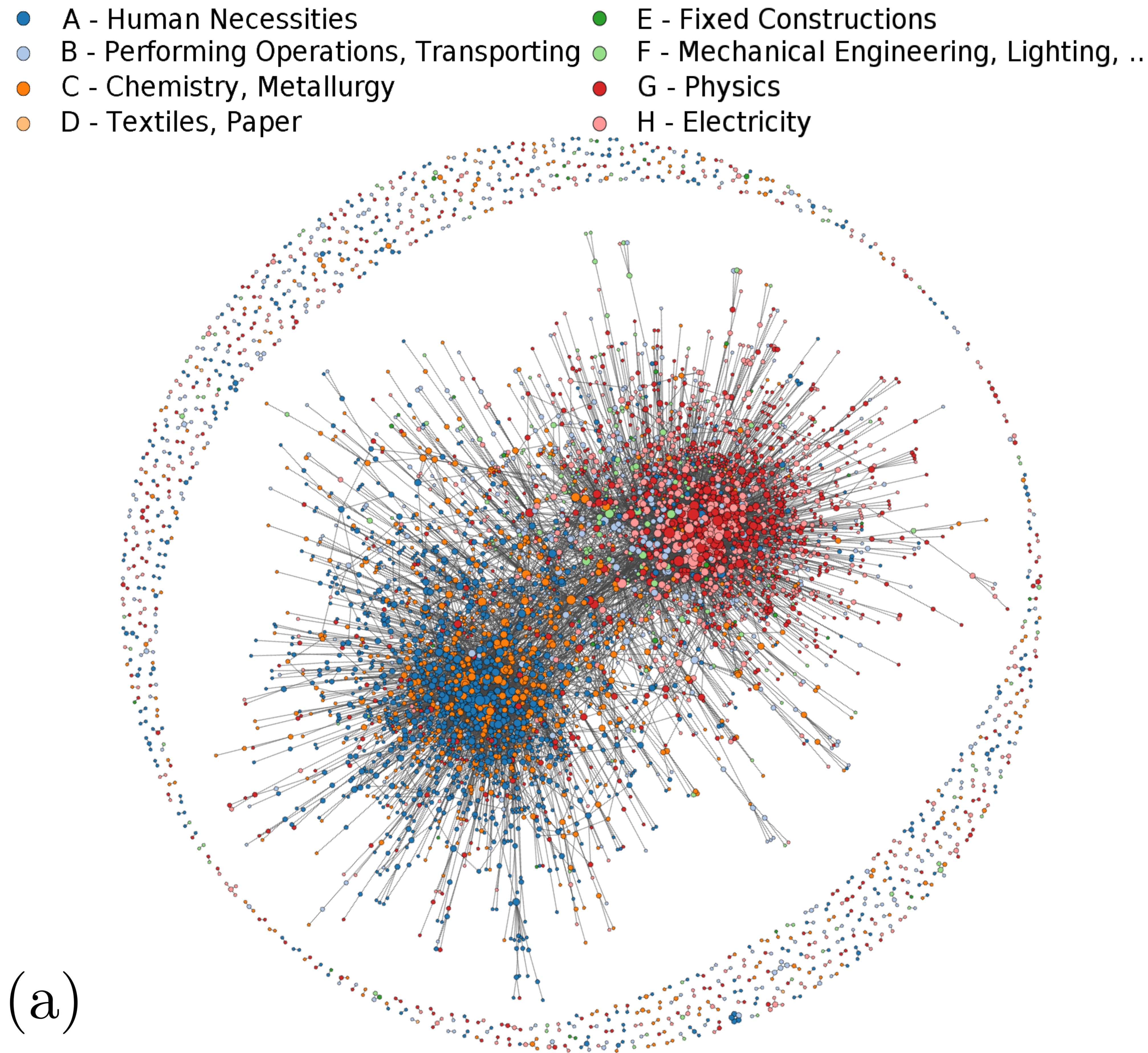}
  \includegraphics[width=0.495\textwidth]{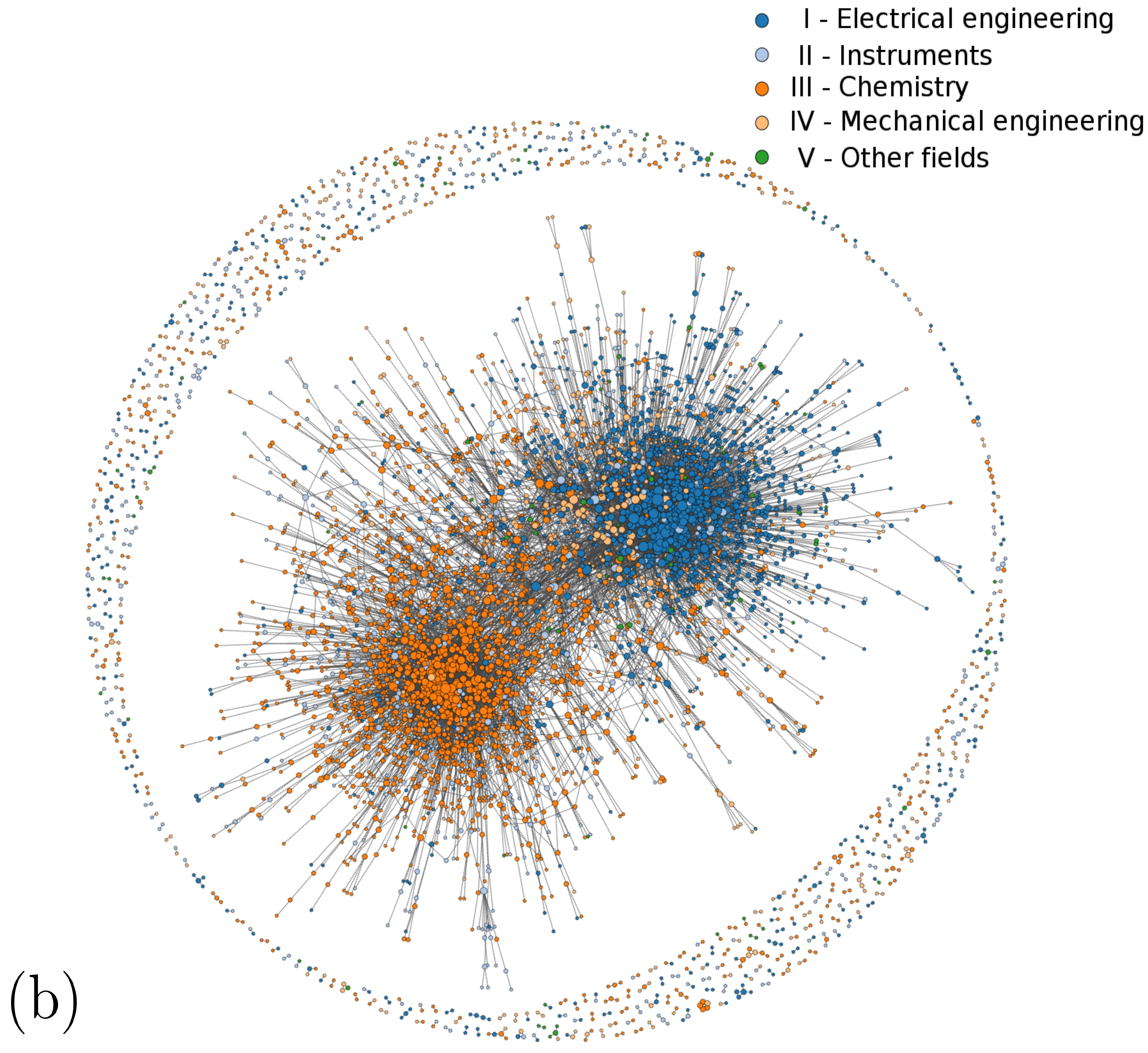}
  \caption{The R\&D network: each node is a firm and its color refers to the domain where the firm has filed more patents between 1984 and 2009. For figure (a) we used the main 8 IPC-sections to classify the patents, while for (b) we used the main 5 areas from ISI-OST-INPI classification scheme. For a discussion about the colors of the nodes see Sect.\ref{sec:firms-posit-knowl}. We use the layout algorithm of \cite{fruchterman_reingold_1991} for both networks. }
  \label{fig:network}
\end{figure}

We define an R\&D network as a set of nodes, or agents (the firms), and links (the alliances between them). 
By R\&D alliance (or collaboration), we refer to an event of partnership between two firms that can span from formal joint ventures to more informal research agreements, specifically aimed at research and development purposes.
To detect such events, we use the SDC Platinum database, provided by \citet{TR-SDC}, that reports all publicly announced alliances, from 1984 to 2009 between several kinds of economic actors (including manufacturing firms, investors, banks and universities). 
In our network representation, we draw an undirected link connecting two nodes every time an alliance between the corresponding firms is announced in the data set. 
When an alliance involves more than two firms (consortium), all the involved firms are connected pairwise, resulting into a fully connected clique.
This procedure is consistent with a previous empirical study \citep{tomasello2016riseandfall}, where there is no conceptual difference between a consortium and a ``standard'' two-partner alliance, which is only a special case of it (and can be thought of as a fully connected clique of size 2).
Fig. \ref{fig:network} shows a visualization of the time aggregated R\&D network, where each node is a firm and links are alliances listed in the above mentioned dataset.  

A quantity that we measure directly from the data prior to the implementation of our agent-based model  is the firms' \textit{activity} distribution.\footnote{For a more detailed definition and more empirical examples on agents' activity in collaboration networks see  \citet{tomasello2014therole} and its Supplementary Information.}
The activity expresses the probability that a firm takes part in any alliance event occurring in a given time window. For the calibration of the present model we use the overall firm activity, measured on the entire observation period of the data set. We define such activity $a_i$ of firm $i$ as the number of alliance events $e_i$ involving firm $i$ divided by the total number of alliance events $E$ involving \textit{any} firm reported in the data set. We then assign such empirical activities $a_i$ to the agents in our computer simulations.


{The SDC Platinum database \citep{TR-SDC} reports approximately 672'000 publicly announced alliances in all countries with a granularity of 1 day.
We apply two filters: first, to select only the alliances characterized by the ``R\&D'' flag; with this, we obtain a list of 14'829 alliances, connecting 14'561 firms.
Second, we keep in our network representation only firms that have a corresponding entry in the patent data set such that we can determine their knowledge positions.}
The patent database used is the Patent Citations Data by the U.S.A. National Bureau of Economic Research (NBER), that contains detailed information on patents granted in the U.S.A. and other contracting countries, from 1971 to present.
Obviously, we select only the entries that have a match with the SDC alliance data set, both with respect to assignees and time period, thus obtaining a total of around 1'400'000 listed patents. Every patent is associated with one or more assignees and with an International Patent Classification (IPC) class. Companies are associated with a unique identifier, and a relatively big part of them (5'168 firms, precisely) are matched to the SDC alliance data set. These firms take part in 7'417 distinct R\&D alliances.

\subsection{Firms positions in knowledge space}
\label{sec:firms-posit-knowl}

\paragraph{Classification schemes}

In this paper we we use -- and compare -- different approaches to determine the knowledge position of a firm. 
Both approaches compute the shares of patents of a firm with respect to two different classification schemes, the Industrial Patent Classification (IPC) and the Fraunhofer ISI, Observatoire des Sciences et des Techniques (OST) and French patent office (INPI) classification (ISI-OST-INPI). 
These classifications differ in the number of classes taken into account, which will correspond to the dimensionality of the knowledge space in which the firms are located. 
IPC operates on  8 dimensions, while ISI-OST-INPI considers 35 dimensions. 
More details are given in the following. 

The IPC, introduced in 1971 by the \textit{Strasbourg Agreement}, is a hierarchical system of symbols for the classification of patents according to the different areas of technology to which they pertain.\footnote{For more information on the International Patent Classification, see \url{http://www.wipo.int/classifications/ipc}.}
A generic IPC category consists of a letter, the so-called ``section symbol'', followed by two digits, the so-called ``class symbol'', and a final letter, the ``subclass''. This four-character term is then followed by a group/subgroup indication, represented by additional digits. A typical IPC term can be written as follows: B34H 6/99.
The sections identified by the IPC are historically stable and amount to 8, from A (human necessities) to H (electricity).
The lower levels are instead subject to more frequent revisions; the eighth and last IPC edition consists of more than 120 classes, 600 subclasses, 7'000 main groups and 60'000 subgroups.

The titles of the 8 sections, as well as a patent count for each section in our data set, is reported in Table \ref{table:comprehensive_model_patent_classes}. We find that the number of patents in all sections reflects their technological dynamism \citep{rosenkopf2007comparing}. 
Indeed, all sections are {not} equally represented. 
For example, the two sections with the lowest patent counts are textiles, paper and fixed constructions, two typical mature industries, {while the sections of Physics and Electricity has the highest patent count. In these sections, patents are often filed by firms belonging to industrial sectors where products innovation and radical innovations play a major role (e.g., from firms working on computer hardware, computer software and electronic components).}

\setlength{\tabcolsep}{6pt}
\renewcommand{\arraystretch}{1.1}

\begin{table}[h]
\footnotesize
\centering
\begin{tabular*}{0.81\linewidth}{ c | l | r }
\hline

\textbf{IPC Section} & \textbf{Title} & \textbf{Patents} \\
\hline

A & Human Necessities & 152,974 \\
B & Performing Operations, Transporting & 244,791 \\
C & Chemistry, Metallurgy & 309,675 \\
D & Textiles, Paper & 12,914 \\
E & Fixed Constructions & 17,842 \\
F & Mechanical Engineering, Lighting, Heating, Weapons & 119,581 \\
G & Physics & 508,815 \\
H & Electricity & 476,437 \\

\hline
\end{tabular*}

\caption[International Patent Classification (IPC) sections and their description.]{International Patent Classification (IPC) sections and their description. The last column reports the number of patents registered in our data set for the corresponding IPC section.}
\label{table:comprehensive_model_patent_classes}
\end{table}

In contrast to the IPC, the ISI-OST-INPI classification scheme is more adapted to the \textit{technological} knowledge space for patents data \citep{schmoch2008concept}. 
As suggested above, this scheme was developed by the Fraunhofer ISI, the Observatoire des Sciences et des Techniques (OST) and French patent office (INPI) in order to overcome problems in the IPC and the US classification scheme. 
There exist various versions of ISI-OST-INPI classification and we chose to use the most updated one, available from PATSTAT, Patent Statistical Database\footnote{\url{https://www.epo.org/searching-for-patents/business/patstat.html}}.
In this version, the scheme groups different IPC codes into 5 technology areas, which are again divided in a total of 35 fields.
The main 5 areas are: 1) Electrical engineering 2) Instruments, 3) Chemistry, 4) Mechanical engineering and 5) Other fields.
In table \ref{tab:ISI-OST-INPI}, we report as an example the classification scheme for the technology area \texttt{Electrical engineering}, as provided from table \texttt{tls901\_techn\_field\_ipc} available in PATSTAT Online, edition Autumn 2016. 
In each entry of the table there is an ISI-OST-INPI code with the corresponding name of the field and IPC codes.
We have created similar tables also for the other four technology areas (not shown).
Using these tables, we assigned to the patents present in our database with one or more IPC codes new ISI-OST-INPI codes.
Our matching procedure was successful since it worked for about 99\% of the patents.

\begin{table}[h]
\centering

\begin{tabular}{cll}
\hline
&\textbf{Electrical engineering} &\\
\hline

1 & \small{Electrical machinery, apparatus, energy} &  \small{F21H, F21K, F21L, F21S, F21V, F21W, F21Y, H01B,}\\
& & \small{H01C, H01F, H01G, H01H, H01J, H01K, H01M, H01R,}\\
&  & \small{H01T, H02B, H02G, H02H, H02J, H02K, H02M, H02N,}\\
& & \small{H02P, H02S, H05B, H05C, H05F, H99Z} \\

2 &\small{Audio-visual technology} &  \small{G09F, G09G, G11B, H04N   3,  H04N   5, H04N   7,}\\
&  & \small{H04N   9, H04N  11,H04N  13, H04N  15, H04N  17,}\\ 
&  &\small{H04N  19, H04N 101, H04R, H04S, H05K }\\

3 &\small{Telecommunications} &  \small{G08C, H01P, H01Q, H04B, H04H, H04J, H04K,} \\
&  &\small{H04M, H04N   1, H04Q} \\

4 &\small{Digital communication}  & \small{H04L, H04N  21, H04W} \\

5 &\small{Basic communication processes} &  \small{H03B, H03C, H03D, H03F, H03G, H03H, H03J,}\\
 & &\small{H03K, H03L, H03M }\\

6 &\small{Computer technology} &  \small{G06C, G06D, G06E, G06F, G06G, G06J, G06K, }\\
&  &\small{G06M, G06N, G06T, G10L, G11C }\\

7 &\small{IT methods for management} &  \small{G06Q }\\

8 &\small{Semiconductors } &  \small{H01L }\\
\hline
\end{tabular}
\caption{ISI-OST-INPI classification scheme based on the IPC, for the technology area of Electrical engineering. The first column  is the ISI-OST-INPI code, the second gives the name of the field and the third column groups the different IPC codes corresponding to the same ISI-OST-INPI code.}
\label{tab:ISI-OST-INPI}
\end{table}

We intend to test our model on a broad set of firms, belonging to several industrial sectors, and therefore exhibiting patent activities distributed across all sections, classes and subclasses. 
For this reason, we have only considered the 8 dimensions (i.e. the first letter) of the IPC code, and the 35 dimensions of the ISI-OST-INPI code. 
Choosing a more refined class- or subclass-level division would result in an excessive patent granularity, meaning a even higher dimensionality for the corresponding knowledge space. 
However, comparing the results for the 8- and the 35-dimensional knowledge space already allows us to draw conclusions about the robustness of our findings with respect to the dimensionality of the knowledge space. 

\paragraph{Knowledge position}

To ensure a match with our model representation, we define the knowledge position of a firm $\mathbf{x}_i \equiv (x_{i1}, x_{i2}, \dots, x_{iD})$ as the set of normalized patent counts $x_{is}$ in each class $s=1,2,\dots D$ (where $D$ is the maximum number of dimensions in the respective classification scheme, i.e. either 8 or 35):
\begin{equation} \label{eq:patent_position}
 x_{is} \equiv \frac{N_{is}}{\sum_s{N_{is}}} \qquad s=1, \dots , D
\end{equation}
$N_{is}$ is the number of patents that the firm $i$ has in a given class $s$. 
In order to compute knowledge distances between pairs of firms, we use the Euclidean metric, similar to \citet{tomasello2016knowledge_exchange}. This means that the knowledge distance between two firms $i$ and $j$ reads as:
\begin{equation} \label{eq:patent_distance}
 \abs{\mathbf{x}_i - \mathbf{x}_j} = \sqrt{\sum_{s=1}^{D}{(x_{is}-x_{js})^2}}
\end{equation}
{In Figs.\ref{fig:network}(a,b) we provide a visualization of the knowledge positions of firms 
using the two patent classification schemes. 
In the time-aggregated R\&D network, nodes represent firms and their colors depend on the patents they have filed between 1984 and 2009. 
In Fig.\ref{fig:network}(a), we have assigned different to each firm the color of IPC-section where it has filed more patents. 
With this, we approximate the knowledge position of each firm for visualization purposes.
In Fig.\ref{fig:network}(b), we apply the same procedure but considering the 5 main areas of the ISI-OST-INPI classification scheme.
From both figures, we find that the two main clusters, which are comprised mainly by pharmaceutical companies (bottom cluster) and firms working on computer hardware, software and communications (top cluster), are dominated by few colors. 
This shows that most alliances occur among firms with a similar knowledge base; alliances with different knowledge bases occur only in specific combinations.}

\paragraph{Distributions of pre-alliance knowledge distances}

Using the definitions provided in Eqs. \ref{eq:patent_position} and \ref{eq:patent_distance}, we can now compute the knowledge positions of the 5'168 firms listed in our data set for the two different classification schemes together with the knowledge position of their alliance partners. 
This allows us to calculate the distribution of the knowledge distances between every pair of allied firms, at the moment of alliance formation (which we know precisely).
We save these pre-alliance distances together with the positions of the firms in knowledge space, to later use this information for setting up the computer simulations. 

\begin{figure}[h]
\begin{center}
(a) 
\includegraphics[width=0.45\textwidth,angle=0]{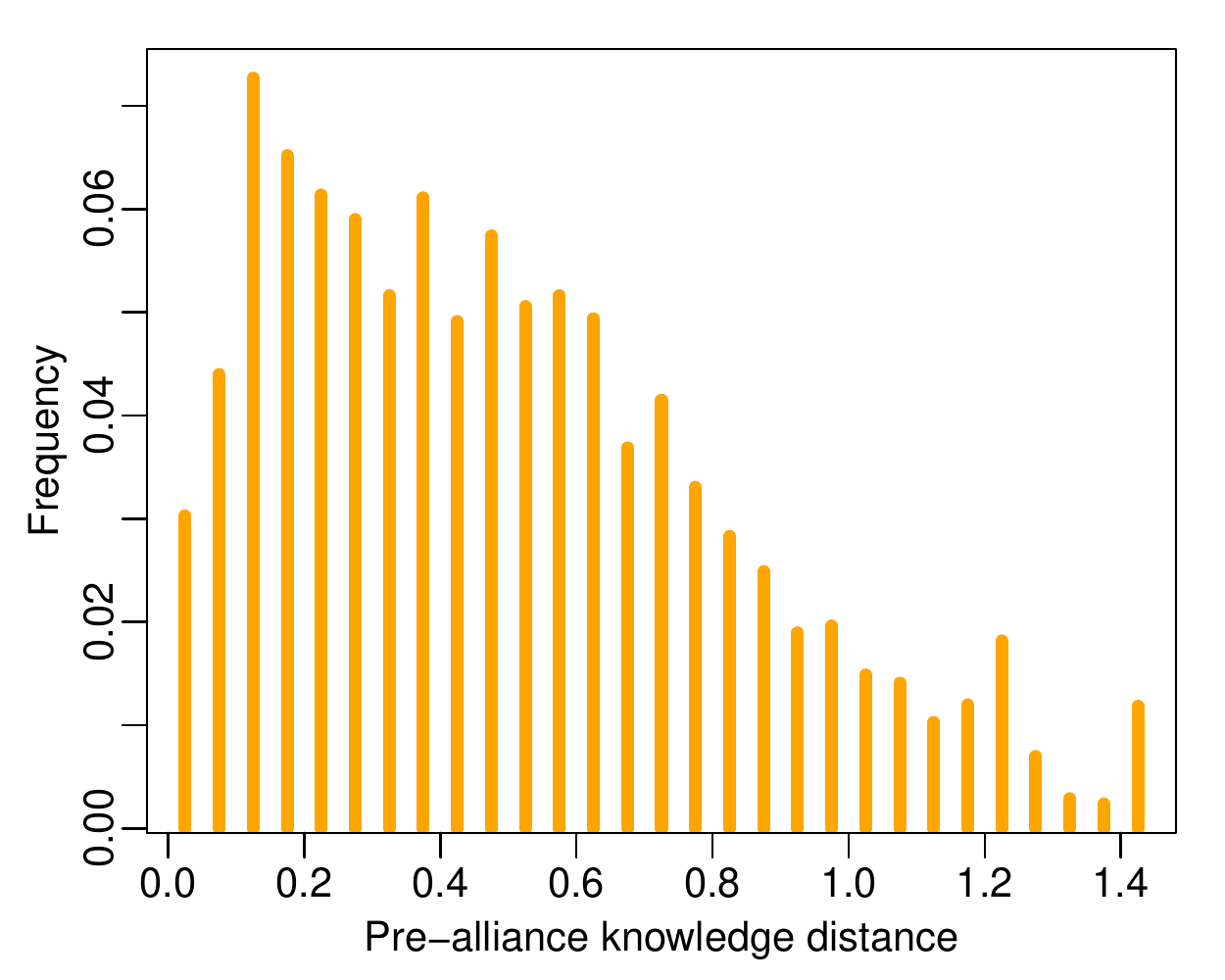}
(b) 
\includegraphics[width=0.45\textwidth,angle=0]{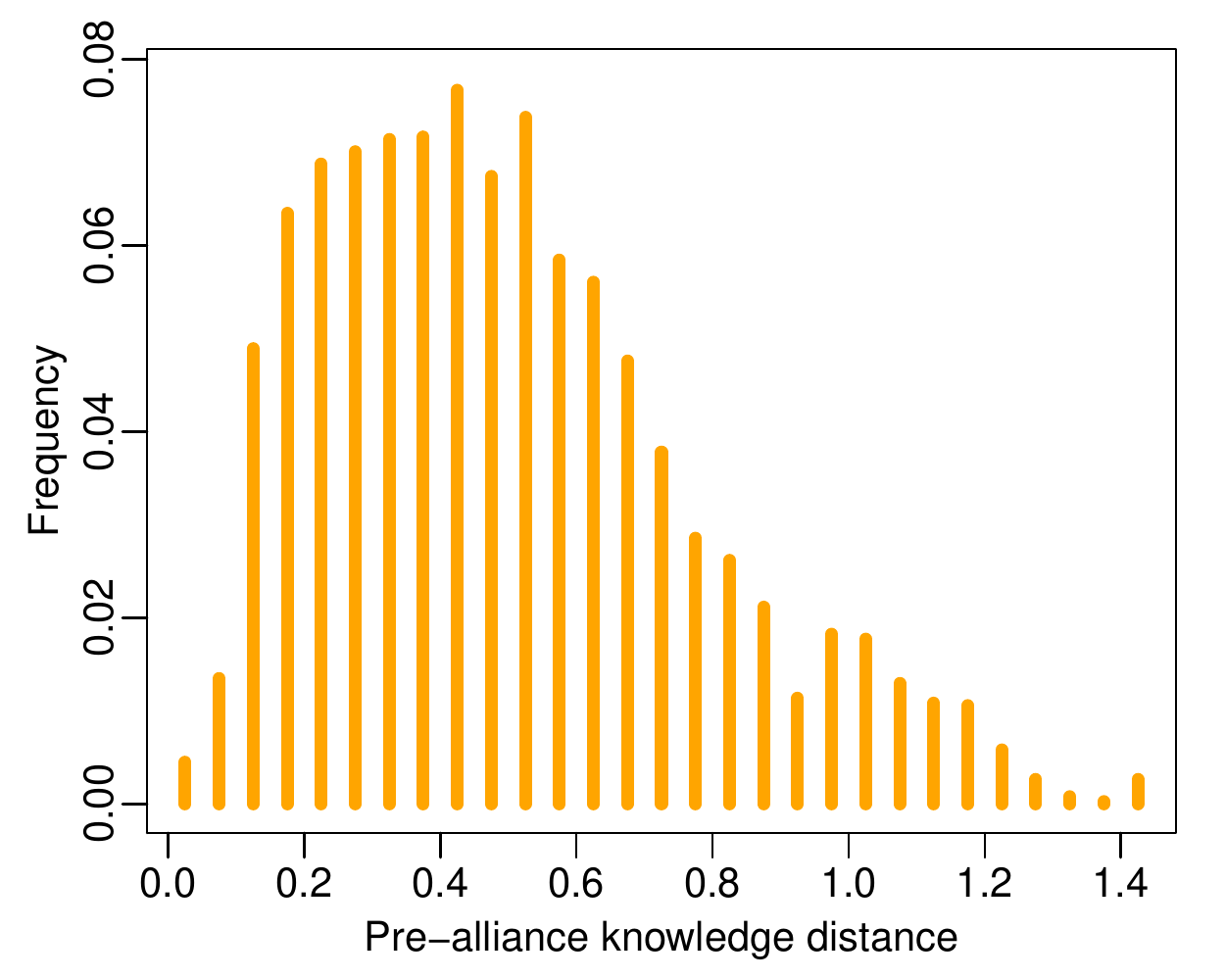}
\end{center}
\caption{Empirical knowledge distance between every pair of partnered firms, as of the day preceding the alliance formation, calculated 
in (a) the 8 dimensional knowledge space defined by the IPC scheme and in (b) the 35 dimensional knowledge space defined by the ISI-OST-INPI classification scheme.}
\label{fig:patent_empirical_prealliance_distr}
\end{figure}

In Fig. \ref{fig:patent_empirical_prealliance_distr} we report the distributions of \emph{pre-alliance knowledge distances} for the two different classification schemes.
The minimum observed value of knowledge distance is 0, while the maximum equals $\sqrt{2}$ (see Eq.\ref{eq:patent_distance}), for normalization reasons.
We find, for both schemes, that the distribution is peaked around an intermediate distance and left-skewed, i.e.~shifted toward small values. Interestingly, we observe that the counts drop when such distances approach zero, meaning that firms with the exact same patenting activity tend not to form alliances.
In addition, it is important to remark that the granularity of the different schemes does not impact the distributions. 

When computing the empirical knowledge position  $\mathbf{x}_i$ of a firm at a given date $t$, we consider all the patents for which the firm has applied in a preceding time window $[t-\Delta t,t]$.
In order to have a reliable and updated measurement, without losing at the same time too much patent information due to a short time window, we use $\Delta t = 5$ years.
We have tested different time windows, ranging from 1 to 10 years, and have found that this only increases the number of missing observations or the noise in the distributions, with no effect on our results.

\section{The model}
\label{sec:model}

We now describe the microscopic interaction rules of our agent-based model. 
In a first phase, the agents form links based on their network features and their social capital; we call this the ``exploration (link formation)'' phase. 
Subsequently, they exchange knowledge through these links, thus approaching each other in a metric knowledge space; we call this ``exploitation (knowledge transfer)'' phase. 
While exchanging knowledge, agents can also form new links; 
in addition, each link can be terminated with a given probability.
Hence, the exploration and exploitation phases are not separated in time.

\subsection{Exploration: link formation}
\label{sec:comprehensive_model_link}

\paragraph{Activation.} We consider a network composed of $N$ agents.
Each agent represents an agent that is endowed with two fundamental attributes, an \textit{activity} and a \textit{label}. 
The activity $a_{i}$ of agent $i$ defines her propensity to engage in a collaboration event. 
We obtain $a_{i}$ from  the distribution of empirical activities extracted from the SDC alliance data set (see Section \ref{sec:data_methods}).
At every time step, agent $i$ initiates an alliance with probability $p_{i}=\eta a_{i} \mathrm{d} t$. 
Consequently, the number of active agents per time step is $N_{A} = \eta \langle a \rangle N \mathrm{d} t$. 
Here $\langle a \rangle$ is the average agent activity and $\eta$ is a rescaling factor that allows to adjust the activation rates.
We fix $\eta=0.0115$ to obtain $N_{A}$ close to $2$ which is the number of active firms per day reported in the alliance data set.
More details will follow on the interpretation of the time step duration $\mathrm{d} t$.

\paragraph{Alliance size.}
Upon activation, an agent selects the number of partners for a collaboration.
We simulate this selection by sampling without replacement a value $n$ from the empirical distribution of alliance sizes, directly measured from the SDC Platinum data set.
With this, we assume that the number of partners, $m=n-1$, with whom the alliance is formed is independent of any characteristic of the active agent.

\paragraph{Label propagation.} 
The second key attribute, called \textit{label}, is used to model
the belonging of firms to communities that are implicitly defined through shared practices and/or behaviors.
In other words, a label can be thought as a membership to a well defined and recognized ``club'' or ``circle of influence''.
We assume that such membership is unique and fixed, i.e. an agent cannot change it nor have more than one.
At the beginning of each simulation all agents are non-labeled.
They can obtain a label in two different ways, (i) by being selected as partner for an alliance or (ii) by initiating one.
In the former case, the non-labeled agent receives the label of the initiator of the alliance, while in the latter she receives a new label that no other agent has in the network.
Both cases are illustrated in Fig. \ref{fig:comprehensive_model_approach}.
It was shown that the described label propagation mechanism can very effectively explain the presence of clusters, or communities, in R\&D networks \citep{tomasello2014therole}.


\paragraph{Selection of the partner categories.} 
The presence of labels allows to distinguish between different types of alliances, dependent on the initiator. 
If the initiator is a labeled agent, she can link to an agent with the same label (with probability $p^{L}_{s}$), to an agent with a different label ($p^{L}_{d}$), or to an agent without a label ($p^{L}_{n}$). 
If the initiator is a non-labeled agent, i.e. she is a \textit{newcomer} in the collaboration network, she can link to a labeled agent (with probability $p^{\mathcal{N}}_{l}$), or to another non-labeled agent ($p^{\mathcal{N}}_{n}$).
The link formation with a labeled agent (described by the probabilities $p^{L}_{s}$,  $p^{L}_{d}$ and $p^{\mathcal{N}}_{l}$) describes \textit{endogenous mechanisms}, because the initiator of the alliance has information about the network position (i.e. social capital) of its potential partners. 
For this case, the two linking probabilities $p^{L}_{s}$ and  $p^{L}_{d}$ allow to tune the importance of the \textit{cohesiveness} as an endogenous driver.
The connection with a non-labeled agent (events $p^{L}_{n}$ and $p^{\mathcal{N}}_{n}$) describes \textit{exogenous mechanisms} because, in this case, the initiator has no information about the social capital of an agent that is not yet part of the network.

\paragraph{Link formation.} 
Once the category (label) of each partner is determined, the initiator of the alliance selects the specific partner. 
To do this, we employ a linear preferential attachment rule, where a agent $j$ is selected with probability proportional to her degree $k_j$ (i.e., the number of previous collaborations with distinct partners).
{This rule is chosen to capture the \textit{prominence} of a firm, namely the history of its previous alliances, as an endogenous driver.
Obviously, this does not apply when the initiator, labeled or not, decides to connect to a non-labeled agent, which has by definition no previous partners ($k_{j}=0$).}
In this case, the partner is selected among all non-labeled agents with equal probability. 
{When the selection process is complete, the initiator connects to its $m$ partners, which accept the offer. 
A variant of the model in which partners can also reject the offer is discussed in \citep{Burkholz2017}.}
In agreement with our representation of the R\&D network, we assume that all the $m$ partners will also link to each other, forming a fully connected clique of size $n=m+1$ with $m (m+1) / 2$ links (see Fig. \ref{fig:comprehensive_model_approach}).

\subsection{Exploitation: knowledge transfer}
\label{sec:comprehensive_model_knowledge}

The second set of microscopic rules models the process of knowledge exchange between pairs of collaborating agents, similar to what has been investigated in \citet{tomasello2016knowledge_exchange}.
Basically, we assume that every agent in the network is located in a metric knowledge space and, as a consequence of its collaborations, approaches its partners in this space. In case of multiple partners, the motion of the focal agent is determined by the vectorial sum of the effects of all of its partners.

\paragraph{Location in a metric knowledge space.}
Here we refer to the description of the (two different) knowledge spaces given in Sect. \ref{sec:firms-posit-knowl}. 
Every agent $i$ ($i=1, \dots , n$) is characterized by a $D$-dimensional vector 
$\mathbf{x}_{i} \equiv (x_{i1},x_{i2}, \dots , x_{iD})$, where the components $x_{i1},x_{i2},...$  are real numbers ranging from 0 to 1. 
In the case of R\&D networks, these numbers are given by the ratios of patents, reflecting the firm's expertise in each of the $D$ dimensions.
Only $D-1$ values of the $x_{is}$ are independent because of the boundary condition that the patent fractions have to sum up to 1.
The dimension of the knowledge space, $D$, is a structural characteristic of the system and fixed  depending on the classification scheme and granularity selected to classify the patents.

\paragraph{Approaching in the metric knowledge space.}
We assume that the existence of a link causes the involved agents to exchange knowledge with their partners {and to align their knowledge bases.}
Hence, as a result of this exchange, they should approach each other in knowledge space.  
To capture this dynamics, every agent is characterized by a \textit{learning rate} $\mu$.  
This parameter is, in first approximation, constant over time and the same for all agents in the collaboration network.
The model dynamics equation can be written as follows:
\begin{equation} \label{eq:comprehensive_model_speed}
  \dot{\mathbf{x}_{i}}(t) = \mu \sum_{j \in \mathcal{N}_{i}(t)}{ [\mathbf{x}_{j}(t)-\mathbf{x}_{i}(t)] }
\end{equation}
where $\mathcal{N}_{i}(t)$ is the set of partners of the agent $i$ at time $t$.
For implementing the model in computer simulations, we use discrete time steps of length $\mathrm{d}t$. The evolution of every agent's position $\mathbf{x_{i}}$ can then be expressed as:
\begin{equation} \label{eq:comprehensive_model_approaching_global}
  \mathbf{x}_{i}(t+ \mathrm{d}t) = \mathbf{x}_{i}(t) + \mu \sum_{j \in \mathcal{N}_{i}(t)}{ [\mathbf{x}_{j}(t)-\mathbf{x}_{i}(t)] } \, \mathrm{d}t
\end{equation} 
{It should be noted from Eq. \eqref{eq:comprehensive_model_speed} that the speed at which a collaborating agent moves in the knowledge space is given by the product of two factors: $\mu$ -- the approach \textit{rate} -- and its distance from the partners.
With this dynamics, the farther agents are in the knowledge space, the faster they move towards each other. 
When the agents' distance decrease, 
the potential for new learning from the collaboration and consequently the approaching speed decrease as well. 
This, eventually, motivates to cancel the collaboration and to terminate the alliance after some time. 
}

Although the dynamics of knowledge exchange is quite simple, it has a number of implications we would like to point out. 
First of all, in the present model proximity in knowledge space is not a  precondition for the agents' interactions.
This is different from other existing models \citep[see, for instance,][]{groeber2009groups, baum2010network, tomasello2016knowledge_exchange} where some sort of ``similarity'' is assumed for a possible collaboration. 
In our model collaboration is determined by the network formation mechanisms, where the different link probabilities are independent of the agents' knowledge positions. 

Second, in our model every link (i.e.~every collaboration) necessarily implies that the involved partners approach each other in the knowledge space. 
This reflects the purpose of the network formation, namely exchange of knowledge. 
Our dynamics assumes that agents approach each other in \emph{all} dimensions of the knowledge space, not just in one particular dimension representing their area of collaboration. 
This reflects the effect of \emph{knowledge spillovers} \citep{baum2010network}, i.e. agents profit from the collaboration not just by the exchange of specific knowledge, but also by learning more general experience.

\paragraph{Alliance termination.}
R\&D alliances have been proven to have a finite duration \citep{phelps2003technological, tomasello2016riseandfall}. 
In order to develop a realistic model, we introduce as a key parameter the characteristic life time $\tau$ of a link. Assuming that the durations of alliances are distributed according to a Poisson process with rate $1/\tau$, the \emph{mean duration} is obviously equal to $\tau$. In our computer simulations, which use discrete time steps of length $\mathrm{d}t$, this translates into the use of a fixed termination probability $p_\mathrm{T} = \mathrm{d} t / \tau $ for any link at any time step.

To keep a simplistic set of rules, we assume that the parameter $\tau$ is a constant, independent of any other feature of the network or the knowledge exchange dynamics or the knowledge stock of the agents. {One possible extension would be to link $\tau$ to the knowledge distance of the two partners, or some other network-related feature.}

\begin{figure}[htpb]
\vspace{6pt}
\begin{center}
 \includegraphics[width=0.8\textwidth]{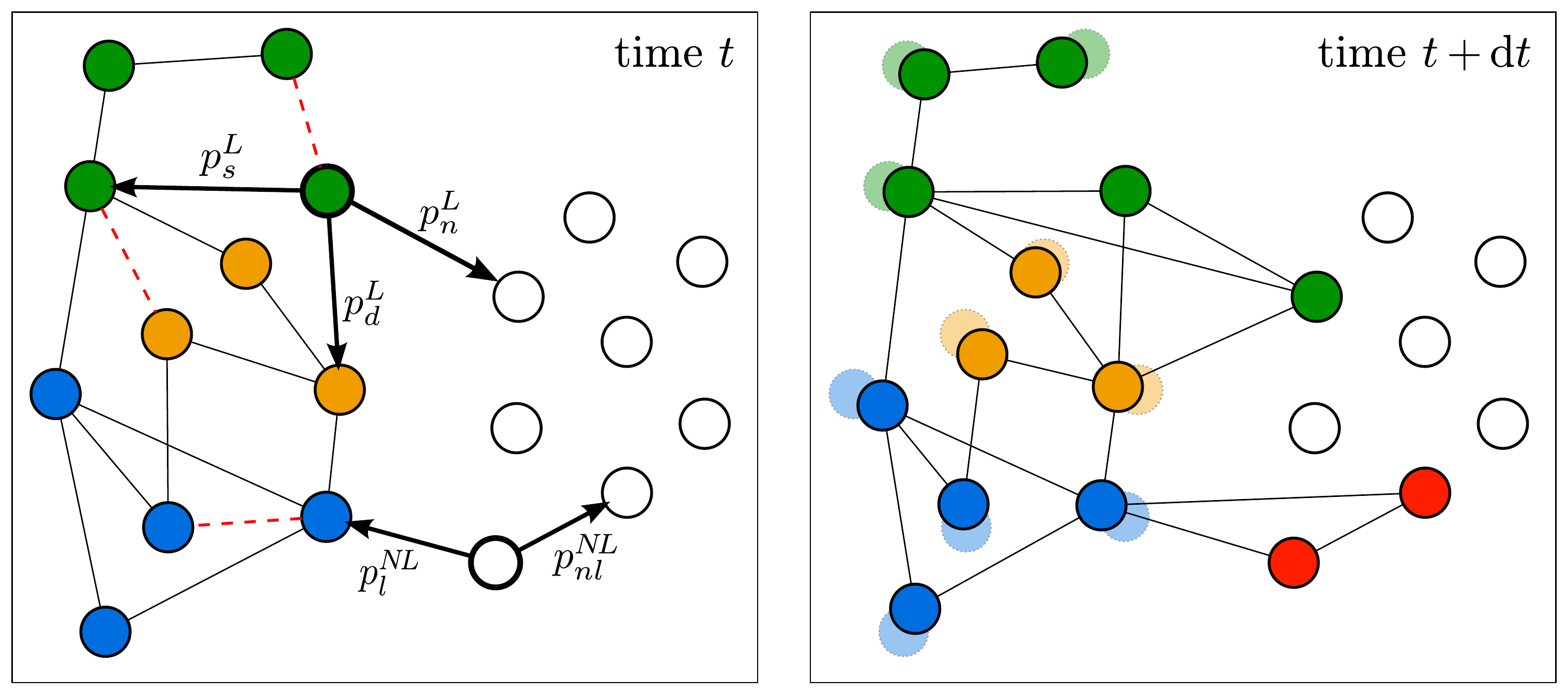}
\vspace{-10pt}
\end{center}
\caption[A representative example of network evolution in a bi-dimensional knowledge space.]{A representative example of network evolution in a bi-dimensional ($D=2$) knowledge space. The position of the agents in the plot corresponds to their coordinates in the knowledge space. At time $t+\mathrm{d}t$, all existing links cause the respective agents to approach in the knowledge space. Furthermore, we illustrate two collaboration events occurring at time $t$. The first one is initiated by a labeled agent (in green), that has linked to $m=3$ new partners, forming a fully connected clique. The second one is initiated by a non-labeled agent, that has linked to $m=2$ new partners and has taken a new arbitrary label (red). At time $t+\mathrm{d}t$, the alliance initiators propagate their labels (respectively, the green one and the red one) to the partners that were not labeled at time $t$ yet. Finally, we illustrate the termination of 3 links (depicted with red dashed lines) at time $t$.}
\label{fig:comprehensive_model_approach}
\end{figure}

To sum up, in this section we have described a set of microscopic rules which aim at reproducing the formation of links in a collaboration network, together with the approach of the agents in an underlying knowledge space.
We summarize the model microscopic rules by means of a visual example in Fig. \ref{fig:comprehensive_model_approach} and report the nomenclature of all parameters in Table \ref{table:comprehensive_model_parameters}.

%
%
%
\setlength{\tabcolsep}{6pt}
\renewcommand{\arraystretch}{1.1}

\begin{table}[h]
\scriptsize
\centering
\begin{tabular*}{0.95\linewidth}{ c | l | l }
\hline

\textbf{Parameter} & \textbf{Explanation} & \textbf{Category} \\
\hline
$p^{L}_{s}$ & Probability that a \boldmath$L$\textbf{abeled} agent chooses an agent with \boldmath$s$\textbf{ame label} & Network formation \\
$p^{L}_{d}$ & Probability that a \boldmath$L$\textbf{abeled} agent chooses an agent with \boldmath$d$\textbf{ifferent label} &  Network formation \\
$p^{\mathcal{N}}_{n}$ & Probability that a \boldmath$\mathcal{N}$\textbf{on-labeled} agent chooses a \boldmath$n$\textbf{on-labeled} agent & Network formation \\
[0.3ex]
$\mu$ & Approaching rate in the knowledge space & Knowledge exchange \\
$\tau$ & Link characteristic life time & Knowledge exchange \\
\hline
\end{tabular*}

\caption[Network formation and knowledge exchange parameters.]{Model parameters and their description. The ``network formation'' parameters are associated with the creation of new links in the collaboration network. The ``knowledge exchange'' parameters are associated with the approach of the agents in a metric knowledge space, occurring as a consequence of a collaboration.}
\label{table:comprehensive_model_parameters}
\end{table}

\section{Model calibration with a two-step procedure}
\label{sec:comprehensive_model_results}

We now calibrate our model against the data, to estimate the value of its parameters. As already mentioned, this is performed in two steps, for network formation and knowledge exchange, by using two data sets, R\&D alliances and patents.

\subsection{Network formation parameters}
\label{sec:network_formation_validation}

In the \textit{first step}, calibrating the network formation model, we fix a set of parameters that we can directly measure from the data, namely the number of agents $N=$5'168, the distribution of the agents activities $a_{i}$, and the distribution of number of partners $m$ per alliance event.

We then estimate the remaining parameters, i.e. $p^{L}_{s}$, $p^{L}_{d}$ and $p^{\mathcal{N}}_{n}$, by running a set of computer simulations, to identify the simulated collaboration network that matches best with the alliance data set.
We stop every computer simulation when the total number of formed alliances equals the number of alliance events reported in the SDC data set, $E=$7'417.
We vary the values of $p^{L}_{s}$, $p^{L}_{d}$ and $p^{\mathcal{N}}_{n}$ in discrete steps spaced by 0.05, in the interval $(0, 1)$.
The parameters $p^{L}_{s}$ and $p^{L}_{d}$ are bounded by the condition $p^{L}_{n} = 1 - p^{L}_{s} - p^{L}_{d} \ge 0$, meaning that their sum has to be smaller or equal to 1. This condition translates into 3'249 points to explore in the 3-dimensional parameter space, for each of which we run 100 simulations (for a total of 324'900 runs).


The networks that we generate by means of computer simulations are matched to the data with respect to three global indicators: average degree $\mean{k}$, average path length $\mean{l}$, and global clustering coefficient $C$,\footnote{For a rigorous definition of these measures, see \citet{tomasello2014therole}.}
For the empirically observed R\&D network, we denote such measures as $\mean{k}^{O\!B\!S}$, $\mean{l}^{O\!B\!S}$ and $C^{O\!B\!S}$, respectively, and their values are $\mean{k}^{O\!B\!S} = 3.45$, $\mean{l}^{O\!B\!S} = 5.05$ and $C^{O\!B\!S} = 0.11$.
\footnote{We find that the present network is slightly denser, more clustered, with a shorter average path length than the R\&D network analyzed in  \citet{tomasello2014therole}. This happens because we now consider only the firms for which patent data are available, not just any firm reported in the SDC alliance data set. These firms typically have more alliance partners than average, thus making the resulting network more dense and connected.}


In order to identify which parameter combination is able to give the best match with the real R\&D network, we use a Maximum Likelihood approach, similar to \citet{tomasello2014therole}. We do not have a set of observations against which we can calibrate our model; instead, we only have one empirical point: the real R\&D network. In particular, we cannot consider the three measures $\mean{k}$, $\mean{l}$ and $C$ as independent, therefore the Likelihood function $\cal L$ reads as:
\begin{equation}
  {\cal L} (p | net^{O\!B\!S}) = f(net^{O\!B\!S} | p)
\label{eq:comprehensive_model_likelihood}
\end{equation}
where $f(\cdot)$ is the joint density function of all parameter combinations $p$ resulting in a network that is equivalent to the observed one, $net^{O\!B\!S}$. Both $p$ and $net^{O\!B\!S}$ are vectors with three components, expressing respectively the three model parameters $p \equiv (p^{L}_{s}, p^{L}_{d}, p^{\mathcal{N}}_{n})$ and the three global network measures $net^{O\!B\!S} \equiv \big( \langle k \rangle^{O\!B\!S}, \langle l \rangle^{O\!B\!S}, C^{O\!B\!S} \big)$. Therefore, we need to find the parameter combination $(p^{L}_{s}, p^{L}_{d}, p^{\mathcal{N}}_{n})$ maximizing the Likelihood ${\cal L} (p | net^{O\!B\!S})$ to generate a network whose macroscopic properties are \textit{sufficiently similar} to the real network $net^{O\!B\!S}$.
By this, we mean that the relative errors from the observed values for the average degree $\epsilon_{\langle k \rangle}$, the average path length $\epsilon_{\langle l \rangle}$ and the global clustering coefficient $\epsilon_{C}$ have to be smaller than a certain threshold $\epsilon^{0}$.

We empirically compute the Likelihood function ${\cal L}$ for each point in the parameter space by counting the fraction of its 100 simulation realizations that fulfill the criteria $\epsilon_{\langle k \rangle} < \epsilon^{0}$; $\epsilon_{\langle l \rangle} < \epsilon^{0}$; $\epsilon_{C} < \epsilon^{0}$. This way, we obtain values that can range from 0 (no realization of that parameter combination fulfills the criteria) to 1 (all of its realizations fulfill the criteria).
For the choice of the error threshold $\epsilon^{0}$, we take a conservative approach and use $\epsilon^{0}=0.02$, that ensures a good matching with the real R\&D network, without cutting out too many points in the parameter space. 



We find that the point in the parameter space with the highest likelihood score has coordinates: $p^{*L}_{s} = 0.45$, $p^{*L}_{d} = 0.2$ and $p^{*\mathcal{N}}_{n} = 0.1$. 
This means that labeled agents show a fairly balanced alliance strategy, with $p^{*L}_{s} = 0.45$, $p^{*L}_{d} = 0.2$, and consequently $p^{*L}_{n} = 0.35$, while non-labeled agents connect rarely with other non-labeled agents ($p^{*\mathcal{N}}_{n} = 0.1$) and prefer to link with labeled ones ($p^{*\mathcal{N}}_{l} = 0.9$). 
In Table \ref{table:comprehensive_model_optimal_parameters}, we report the full set of parameter values maximizing the likelihood score, together with the values of average degree, average path length and global clustering coefficient for the simulated and the real R\&D networks.


\begin{table}
\footnotesize
\centering
\begin{tabular*}{0.95\linewidth}{|@{\extracolsep{\fill} }c | c || c | c || c | c |}
\hline
\multicolumn{4}{| c ||}{\textbf{Optimal simulated R\&D network}} & \multicolumn{2}{ c |}{\textbf{Real R\&D network (with patents)}} \\
\hline
Model parameter & Value & Measure & Value & Measure & Value \\
\hline
$p^{*L}_{s}$ & 0.45 & $\langle k \rangle^{*}$ & $3.48 \pm 0.01$ & $\langle k \rangle^{O\!B\!S}$ & 3.45 \\
[0.5ex]
$p^{*L}_{d}$ & 0.2 & $\langle l \rangle^{*}$ & $5.02 \pm 0.08$ & $\langle l \rangle^{O\!B\!S}$ & 5.05 \\
[0.5ex]
$p^{*L}_{n}$ & 0.35 & $C^{*}$ & $0.111 \pm 0.007$ & $C^{O\!B\!S}$ & 0.109 \\ 
[0.5ex]
$p^{*\mathcal{N}}_{n}$ & 0.1  &  &  &  &  \\
[0.5ex]
$p^{*\mathcal{N}}_{l}$ & 0.9  &  &  &  &  \\
[0.3ex]
\hline
\end{tabular*}

\caption[Model parameter set $p^{*}$ defining the optimal simulated R\&D network.]{Link formation parameters $p^{*}$ defining the optimal simulated R\&D network. The average degree, average path length and global clustering coefficient of the 100 realizations of the optimal R\&D network are compared to their empirical counterparts.}
\label{table:comprehensive_model_optimal_parameters}
\end{table}

These results are in line with those presented by \citet{tomasello2014therole}. 
However, the R\&D network with patent data, used here, exhibits an even stronger tendency to favor connections with labeled agents (i.e.~incumbent firms) than the pooled R\&D network including all firms, irrespectively of their patenting activity. Let us spend a few words on the comparison between these two networks.

Due to the fact that our analysis in now restricted only to firms for which patent data are available, one could expect either an increase in the importance of network endogenous mechanisms, given that we are considering, on the one hand, larger and more active firms -- or an increase in the importance of exogenous mechanisms, given that we are considering, on the other hand, firms for which the technological dimension could be more relevant in the alliance formation strategy.
Our data confirm the first hypothesis, that is the increase in the relevance of network endogenous mechanisms, which results in higher probabilities for the agents to collaborate with agents that are already part of the network, and therefore already labeled. This behavior is present irrespective of whether the alliance event is initiated by a labeled or a non-labeled agent: precisely, 65\% of the collaborations initiated by labeled agents ($p^{*L}_{s}+p^{*L}_{d} $), as well as 90\% of the collaborations initiated by non-labeled agents ($p^{*\mathcal{N}}_{l}$), involve a labeled agent as a partner.

\subsection{Knowledge exchange parameters}  
\label{sec:knowledge_exchange_validation}

In the \textit{second step}, we fix the network formation parameters to the values obtained in the first step, and run a second set of computer simulations. 
This time we estimate the knowledge exchange parameters, i.e. $\mu$ and $\tau$, by identifying the simulated collaboration network that best matches with the patent data set.
To quantify the knowledge space, we use either the eight main sections of the IPC scheme {or the 35 technological fields of the ISI-OST-INPI classification scheme}, i.e. the dimensions are set to $D=8$ or $D=35$.

\paragraph{Pre-alliance conditions}

In order to calibrate the dynamics of knowledge transfer, we need to assign to the agents a current position in the respective knowledge space, to calculate their future positions. 
Following the model of network formation, we need to distinguish between the agent that initiates a collaboration (when becoming active), and the $m$ collaborators chosen by the initiator. 

A naive approach would assume that we first randomly choose an initiator with its initial position in knowledge space, then randomly choose $m$ collaborators, their distances in knowledge space randomly sampled from the empirical distribution of pre-alliance distances shown in Figure \ref{fig:patent_empirical_prealliance_distr}.  
Second, we run the knowledge exchange dynamics of Eq. \eqref{eq:patent_distance}, to calculate the expected movement in knowledge space for a given set of parameters $\tau$, $\mu$. 
Eventually, we compare the distribution of distances for various $\tau$, $\mu$ with the empirical distribution of post-alliance knowledge distances, to find out which set of parameters matches best. 

\begin{figure}[h]
\begin{center}
(a)
\includegraphics[width=0.45\textwidth,angle=0]{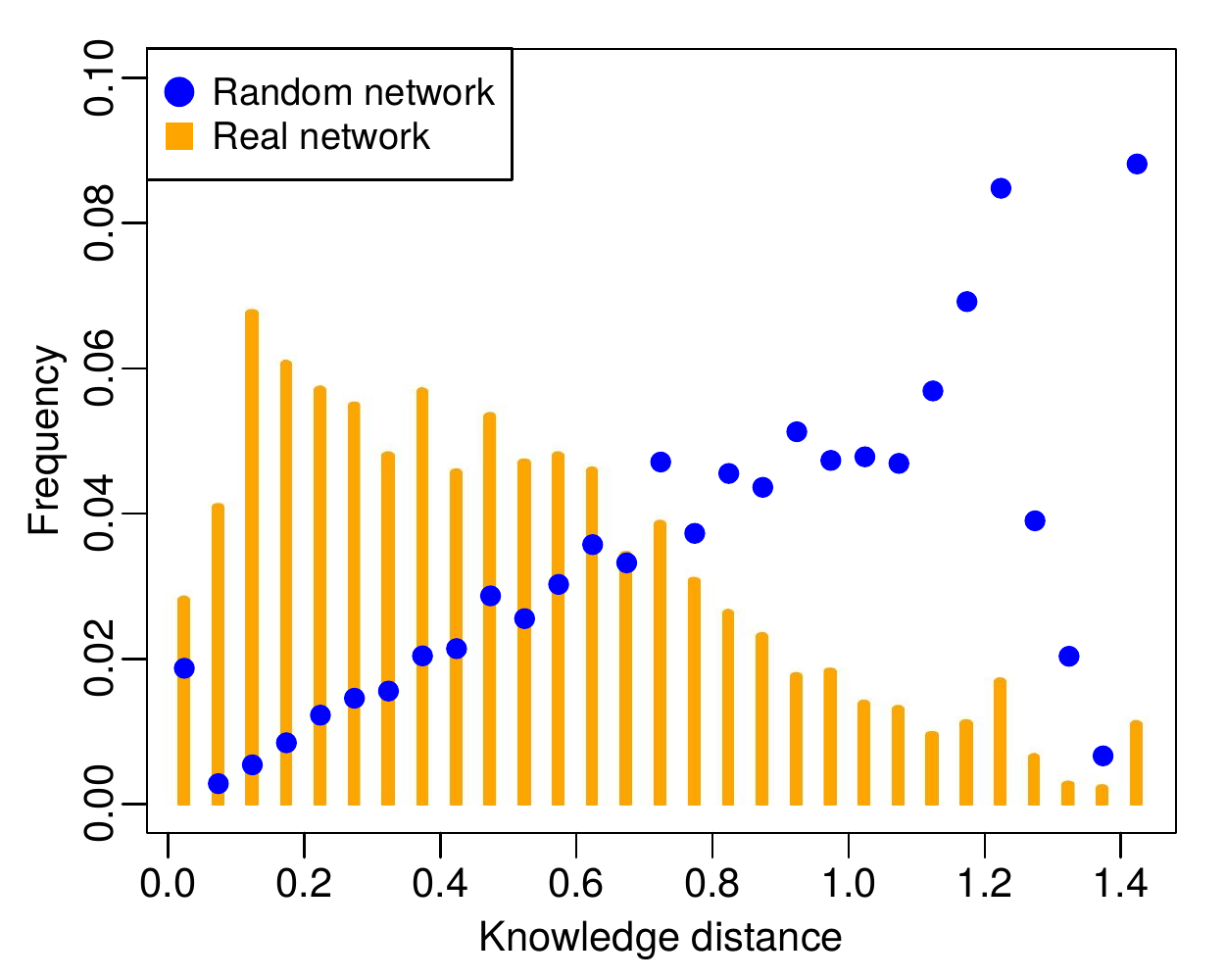}
(b)
\includegraphics[width=0.45\textwidth,angle=0]{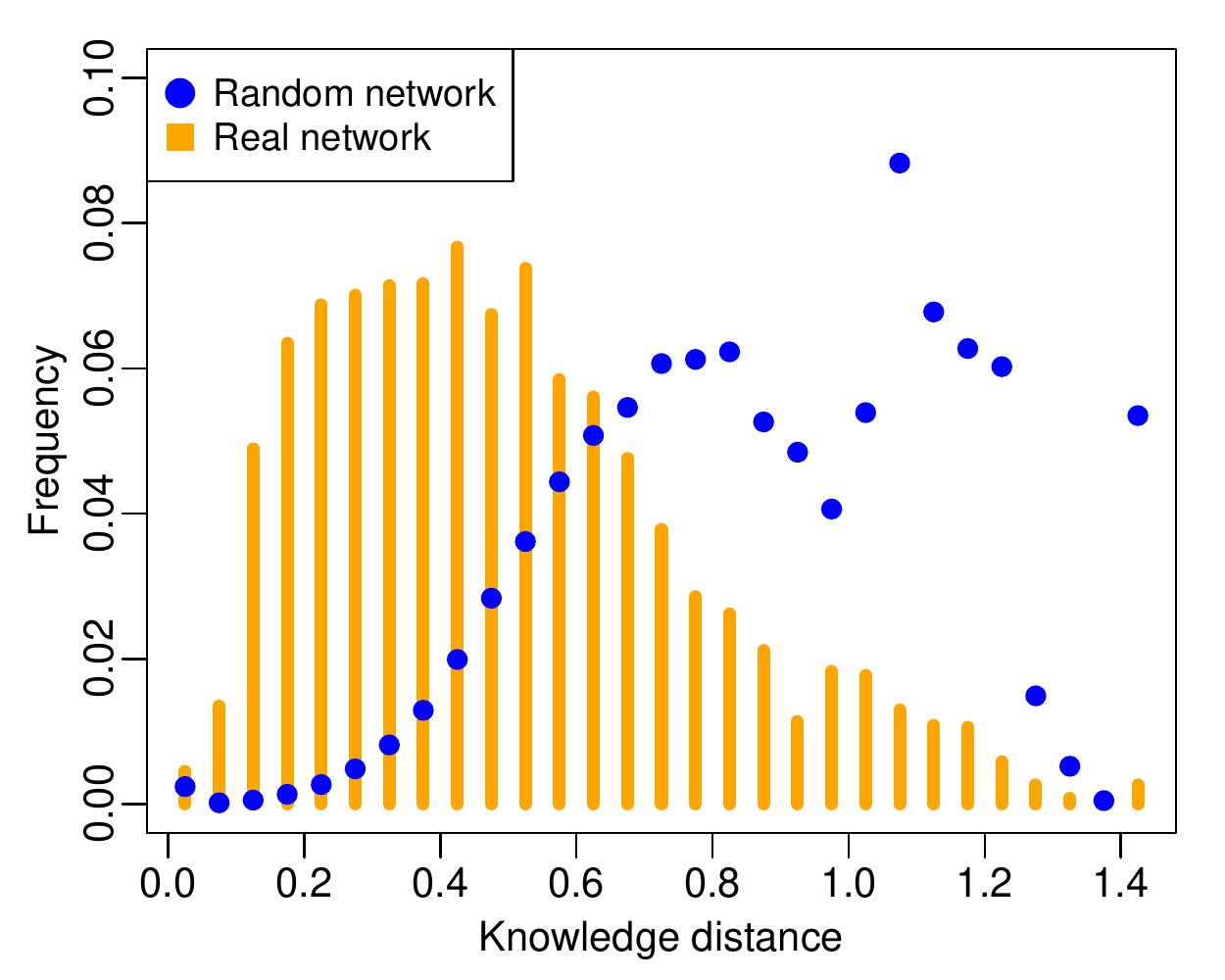}
\end{center}
\caption{Pre-alliance distance distributions from the empirical and a randomized R\&D network. In (a) we used the IPC scheme to calculate the firms positions, while in 
(b) the ISI-OST-INPI scheme.}
\label{fig:emp-sim-pre}
\end{figure}
While the second part of the procedure is correct, the first part is based on the wrong assumption that firms \emph{randomly} choose their collaboration partners from the knowledge space. 
Figure \ref{fig:emp-sim-pre} shows, for the two different knowledge metrics used, how the distribution of pre-alliance distances should look like if every possible knowledge distance would be realized. 
We note the strong deviations between the random and the empirical distributions.
First, the random distributions appear right-skewed while the empirical are left-skewed. 
Second, the average pre-alliance distance are around $0.9$ in the random case, while the averages of the empirical pre-alliance distances is much smaller, around $0.6$.

With this, we can conclude that the empirical pre-alliance distance distributions cannot be explained by assuming that firms create alliances without considering the position of their possible collaborators in the knowledge space. 
Hence, we need to essentially consider the \emph{full} agent-based model -- not to calibrate the dynamics of knowledge exchange, but to correctly determine the \emph{initial conditions} for the knowledge exchange dynamics. 
This lends strong support to consider the \emph{combined processes} of network formation and knowledge exchange, as it is proposed in our model, instead of investigating knowledge exchange in isolation. 
 

In order to determine the pre-alliance conditions in knowledge space for our model at a given time $t$, we distinguish between agents that are \emph{not} currently, at time $t$, involved in any collaboration  and those that \emph{are} currently involved. 
Agents that \emph{are} involved, already have a position in knowledge space that reflects their previous interaction with other agents during the simulation up to time $t$.  
Thus, we decide to keep these (simulated) positions at time $t$ as starting point for their knowledge exchange in the new alliance. 
For those agents that are \emph{not} involved in a collaboration at time $t$, we obtain the initial conditions from sampling from the empirical data.
Precisely, the position of an initiator that is not currently involved in an alliance is sampled from the  distribution of pre-alliance positions obtained from the real patent data. 
And for the collaborating agents that are \emph{not} involved in any other alliance at time $t$, we assign a \emph{knowledge distance} by sampling with replacement from the empirical distribution of pre-alliance distances given in Figure \ref{fig:patent_empirical_prealliance_distr}. 

This procedure of determining the pre-alliance distance distribution mixes up two conceptually different information.
Part of it is obtained from \emph{simulations}, this way taking into account the path dependence of the recent history in collaborations, i.e. the active partners in alliances and their influence on knowledge exchange.
Another part of information comes from the \emph{empirical distribution} of pre-alliance knowledge positions/distances that reflects e.g. preferences of agents in choosing partners at shorter distances.    
Further, it captures the fact that firms not engaged in any R\&D alliance can still perform related activities and thus move in knowledge space, which is reflected by their new position assigned when engaging in  a new alliance. 
We emphasize again that, without the empirical information, we would randomly pair agents that likely had not chosen to collaborate or we would assume that agents do not move in knowledge space by themselves. 
Without the simulations, on the other hand, we would create problematic artifacts in all cases where agents already involved in a collaboration are chosen to participate in a new alliance. 
In such cases, we cannot assign two positions in knowledge space to the same agent or randomly switch between profiles. 
Thus, the best solution is to keep the evolution of agents \emph{during} existing collaborations into account, as a precondition for new ones. 

\begin{figure}[h]
\begin{center}
(a)
\includegraphics[width=0.45\textwidth,angle=0]{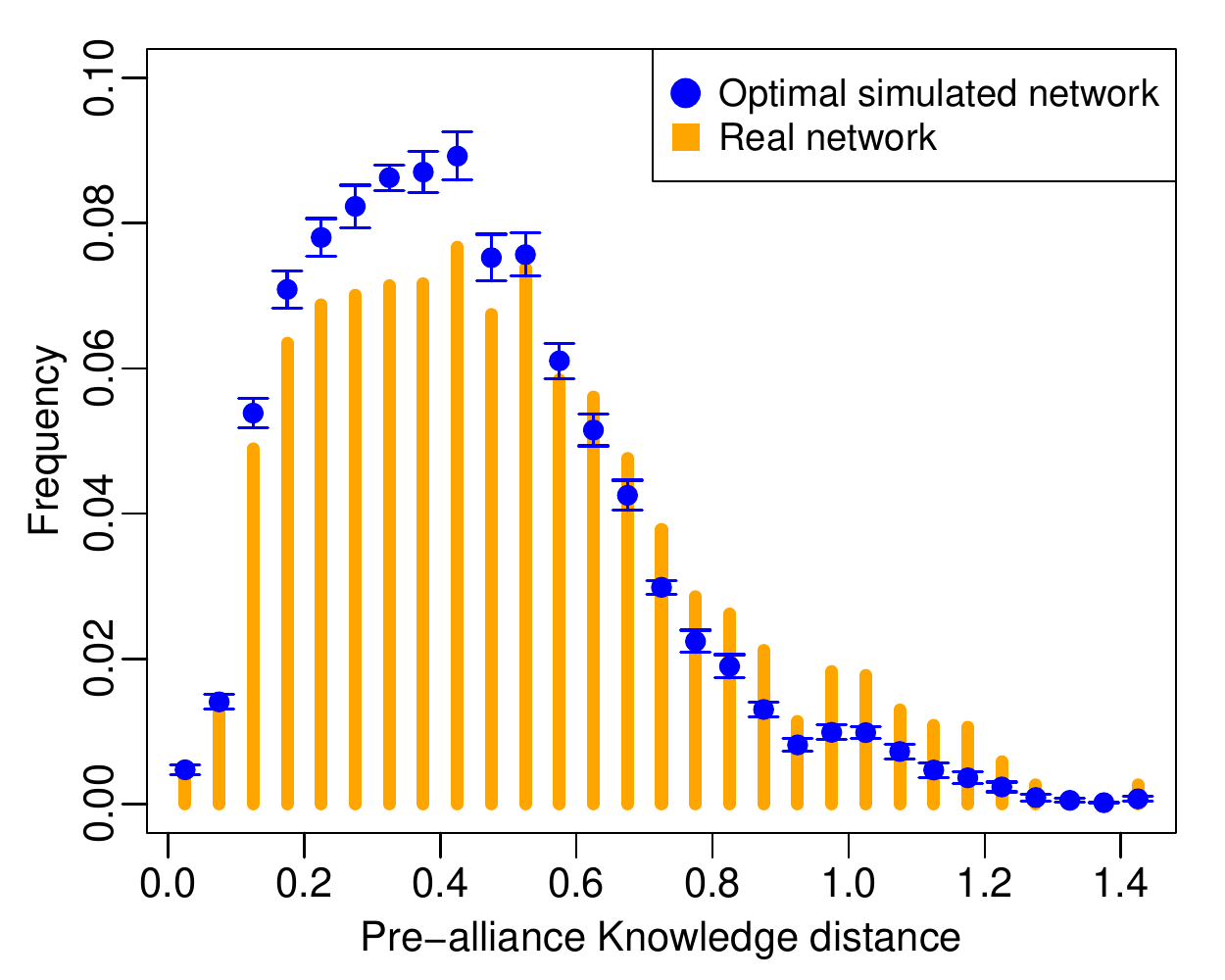}
(b)
\includegraphics[width=0.45\textwidth,angle=0]{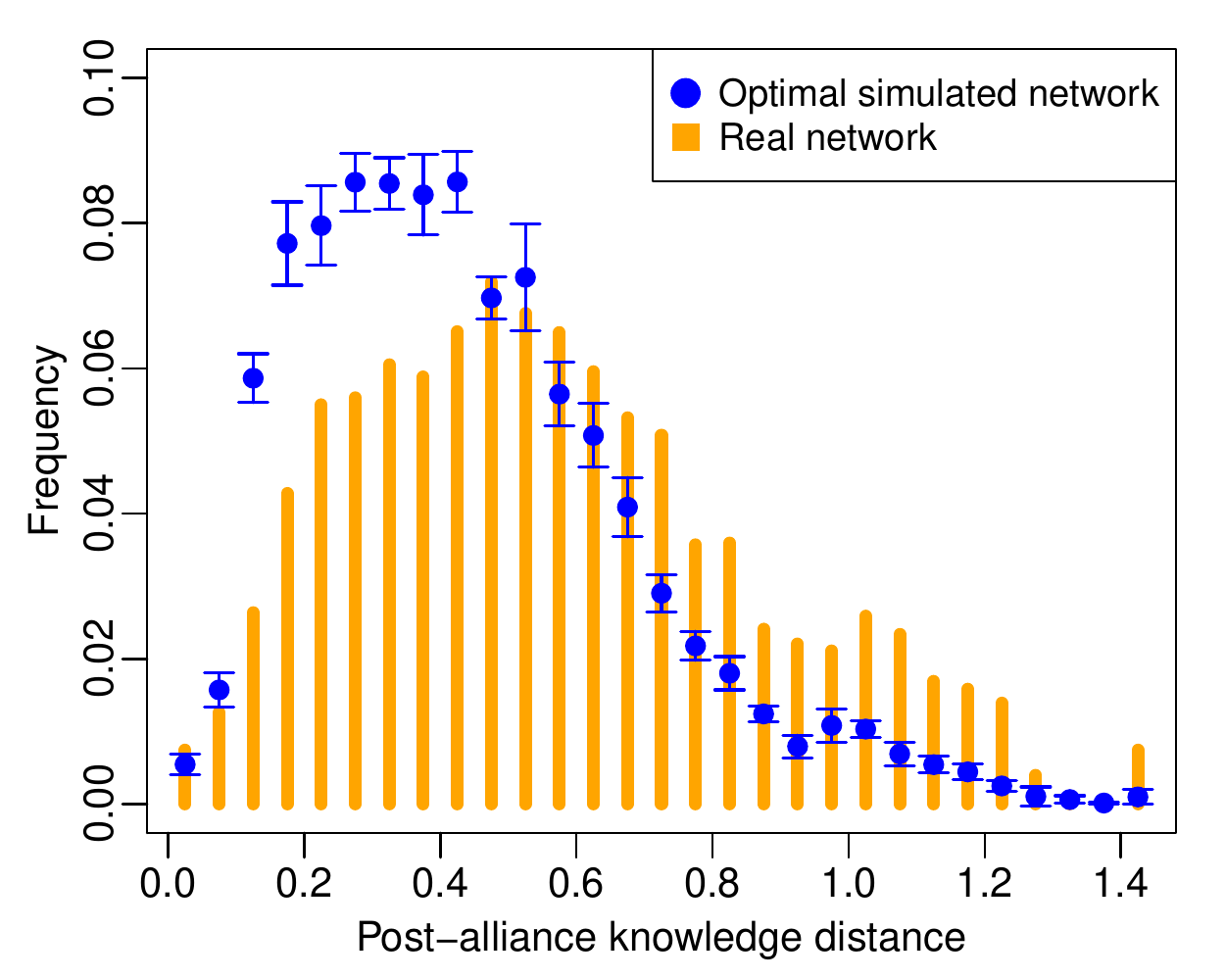}
\end{center}
\caption[Empirical and simulated pre-alliance knowledge distances.]{Empirical and simulated distances between firms at the moment of alliance formation and at the assumed termination of alliances after $\tau=700$ days.
In both plots the distances are calculated in the 35 dimensional space defined by the ISI-OST-INPI classification, the blue circles correspond to the mean values and the error bars correspond to the standard deviations of all the measures we study on the 100 realizations of the optimal simulated R\&D network.}
\label{fig:simul_pre_distances}
\end{figure}

This leads us to an important question that we need to answer before we can discuss the details of the parameter calibration: What is the error that we may introduce by mixing these two source of information for determining the initial conditions? 
In Fig. \ref{fig:simul_pre_distances}(a), we show the distribution of pre-alliance distances that follows from the constraint of respecting current knowledge positions in comparison to the empirical distribution.
We find that the simulated distribution matches the empirical one over a large range; however, the simulations 
overestimate the probability of having alliances among firms separated by a small knowledge distance. 
This deviation is significant only in the range of distances between 0.2 and 0.4, where the distribution has its maximum. 

Obviously, such deviations in the initial conditions are amplified during the simulated knowledge exchange, as can be seen in Fig. \ref{fig:simul_pre_distances}(b) which shows the \emph{post-alliance distance distribution}. 
Precisely, compared to the empirical distribution of \emph{pre}-alliance distances, in the empirical distribution of \emph{post}-alliance distances the probability to have a small knowledge distances has decreased, whereas it has increased in the corresponding simulations. 
We will comment on this interesting observation further in Sect. \ref{sec:conclusions}.

{At this point, we just emphasize that the empirical distribution of pre-alliance distances is much better matched by the distribution obtained from our simulations that use the selection process described above (see Fig. \ref{fig:simul_pre_distances}(a)) compared to the distribution obtained assuming a random selection process (see Fig. \ref{fig:emp-sim-pre}(b)).
Indeed, when we perform a two-sided Kolmogorov-Smirnov (KS) test between our simulated distribution of pre-alliance distances and the empirical one, we find an average $\mathcal{D}$-statistic 10 times smaller, i.e. better, compared to the $\mathcal{D}$-statistic coming from the KS-test performed between the distributions  shown in Fig. \ref{fig:emp-sim-pre}(b).
We disregard the $p$-value of the KS-test, because we are not interested in statistically inferring the provenience of the two distributions from a hypothetical common distribution.
Our aim is instead to quantify the similarity between pairs of distributions, a measure that is already fully captured by the $\mathcal{D}$-statistics of a two-sided KS-test.}
Hence, in the following we will take the distribution of pre-alliance distances shown in Fig. \ref{fig:simul_pre_distances}(a) as a good proxy for the initial condition at the moment of alliance formation.


\paragraph{Optimal parameters}

In the subsequent computer simulations we vary the values of the two remaining knowledge exchange parameters, i.e. the agents' approaching rate $\mu$ and the characteristic alliance life time $\tau$. 
We consider the values $5\times10^{-8}$, $10^{-7}$, $5\times10^{-7}$, $10^{-6}$, $5\times10^{-6}$, $10^{-5}$ for the parameter $\mu$ and the values  700, 1000, 1500 and 2000 for the parameter $\tau$, thus having a total of 24 points to explore in the parameter space. 
The interpretation of the parameter $\tau$ is straightforward: as explained in Section \ref{sec:comprehensive_model_link}, we adjust the activation rate of the agents such that the length of a time step $\mathrm{d}t$ can be directly interpreted as 1 day. Therefore, the value of $\tau$, which is by design expressed in time steps, can be thought of as the characteristic duration of a real alliance in days.

For each of the 24 parameter combinations, we run 100 simulations that combine the network formation process (using the optimal parameters determined) and the knowledge exchange dynamics.
This results in a total of 2'400 simulation runs only to complete the second step of our calibration procedure, namely to determine the optimal knowledge exchange parameters. 
We store the distributions of post-alliance knowledge distances and knowledge distance shifts in each run. Similar to the first step, we stop every computer simulation when the total number of collaborations equals the number of alliance events reported in the SDC data set, $E=$7'417.
 
As explained, the distribution of pre-alliance distances shown in Fig. \ref{fig:simul_pre_distances}(a) is used as an input of the simulations. 
Thus, we use the distribution of post-alliance knowledge distances, obtained from each of the 100 simulations for every parameter combination, to compare it to the respective distance distribution obtained from the empirical R\&D network. 
This comparison relies on determining the \emph{post-alliance} time. 
It becomes a problem for the empirical data because the termination dates of  alliances are  not available.
In the simulations, however, we have assumed that alliances have a duration $\tau$ and are terminated stochastically, afterwards.
To allow for comparison, we compute, from the empirical data, the knowledge distance between every pair of linked firms after the same time period $\tau$, in days, as used in the corresponding simulation.

To compare the two distributions of simulated and empirical knowledge distances, we use the two-sided KS-test that assigns a score, the $\mathcal{D}-$statistics, to each simulated distribution. 
The value of the $\mathcal{D}-$statistics decreases as the simulated and the empirical distributions become more similar, hence, it is used here as goodness score for each simulation. 
We finally average the 100 score values for the 100 simulations, for each combination of the parameters. 

\begin{figure}[htbp]
\begin{center}
\includegraphics[width=0.45\textwidth,angle=0]{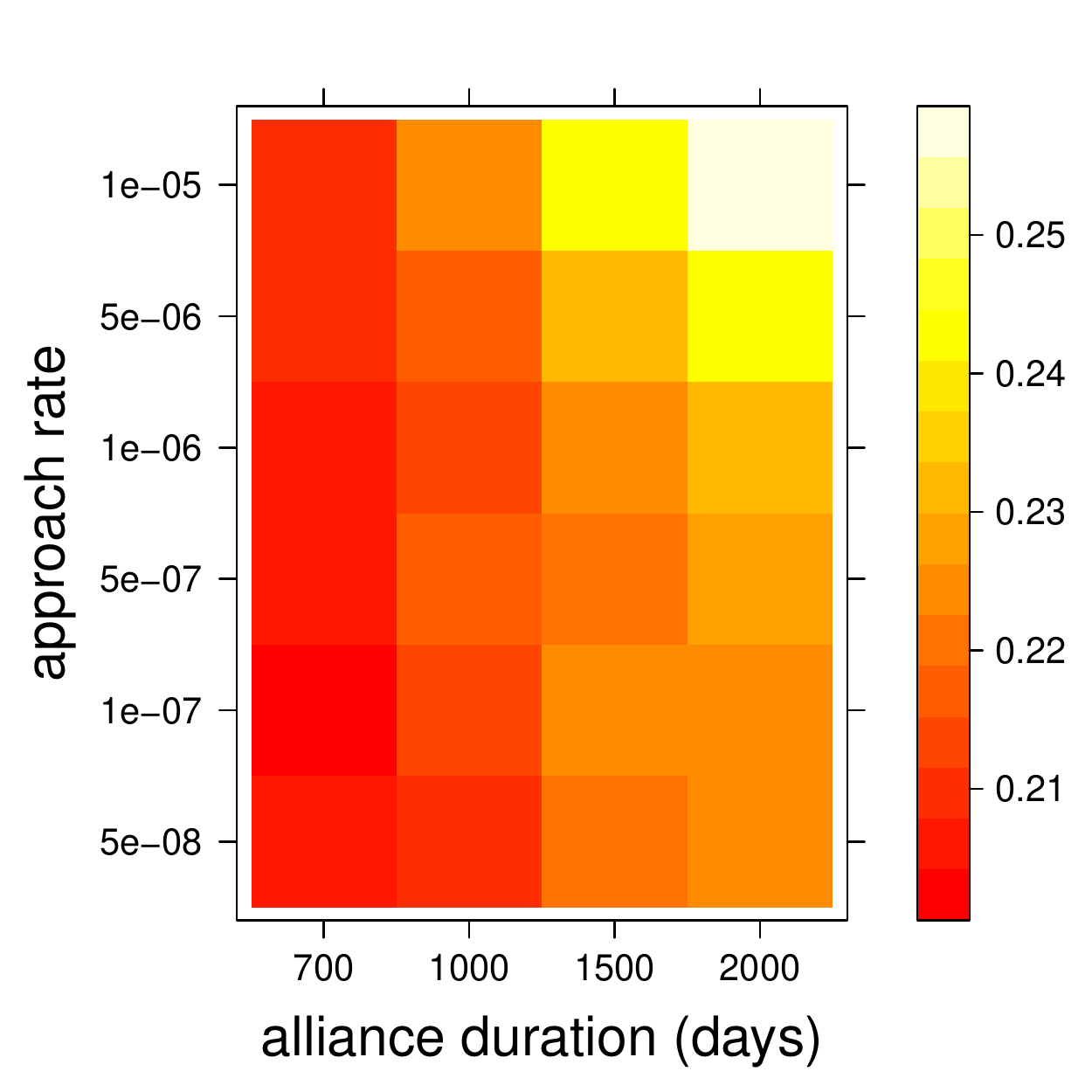}
\end{center}
\caption[Goodness score for every point in the parameter space, depicted by means of a heat-map.]{Goodness score for every point in the parameter space, depicted by means of a heat-map. The color scale corresponds to the score value; the lower the score, the closer the simulated distribution of post-alliance distances is to the empirical one. The simulations and the distances have been obtained considering the 35 dimensional space defined by the ISI-OST-INPI classification scheme.}
\label{fig:comprehensive_model_goodness}
\end{figure}

The resulting goodness scores are presented in the heat map plot of Fig. \ref{fig:comprehensive_model_goodness}.
It shows the bi-dimensional parameter space of alliance duration $\tau$ and learning rate $\mu$.
As the color code indicates, we find an entire region of parameters with maximized goodness score
for parameter combinations with medium to large $\mu$, but low $\tau$ values. 

Although many parameter combinations exhibit a similar, low goodness score, i.e.~they are fairly equally able to reproduce the empirical post-alliance knowledge distance distribution, the best parameter sets can be ranked quantitatively. 
We find that the parameter point yielding the best goodness score is identified by the following coordinates: {$\mu=10^{-7}$} and $\tau=700$. This means the optimal simulated collaboration network exhibits a low approaching rate, and a characteristic alliance duration slightly shorter than 2 years.
This is not only consistent with previous theoretical and empirical observations \citep{phelps2003technological, inkpen01:_why_do_some_strat_allian}, but also in line with our previous assumption \citet{tomasello2016riseandfall} to terminate alliances after 3 years in the empirical network representation. 
Taking into account that we have obtained this result here by using two different data sets and an involved
agent-based model, the agreement is even more remarkable. 

\subsection{Robustness analysis}
\label{sec:results}

\paragraph{Distribution of post-alliance knowledge distances}

Already for the model calibration, we addressed the problem that the exact durations of R\&D alliances are not known from the data set. 
This leads to the above estimations of the optimal duration $\tau$ conditional on the knowledge transfer rate $\mu$. 
However, we can also independently investigate how sensitive the distribution of post-alliance distances responds to changes of the (unknown) duration of alliances. 
This is done in the following two steps for both of the knowledge space metrics used.   

\begin{figure}[h]
\begin{center}
(a)
\includegraphics[width=0.45\textwidth,angle=0]{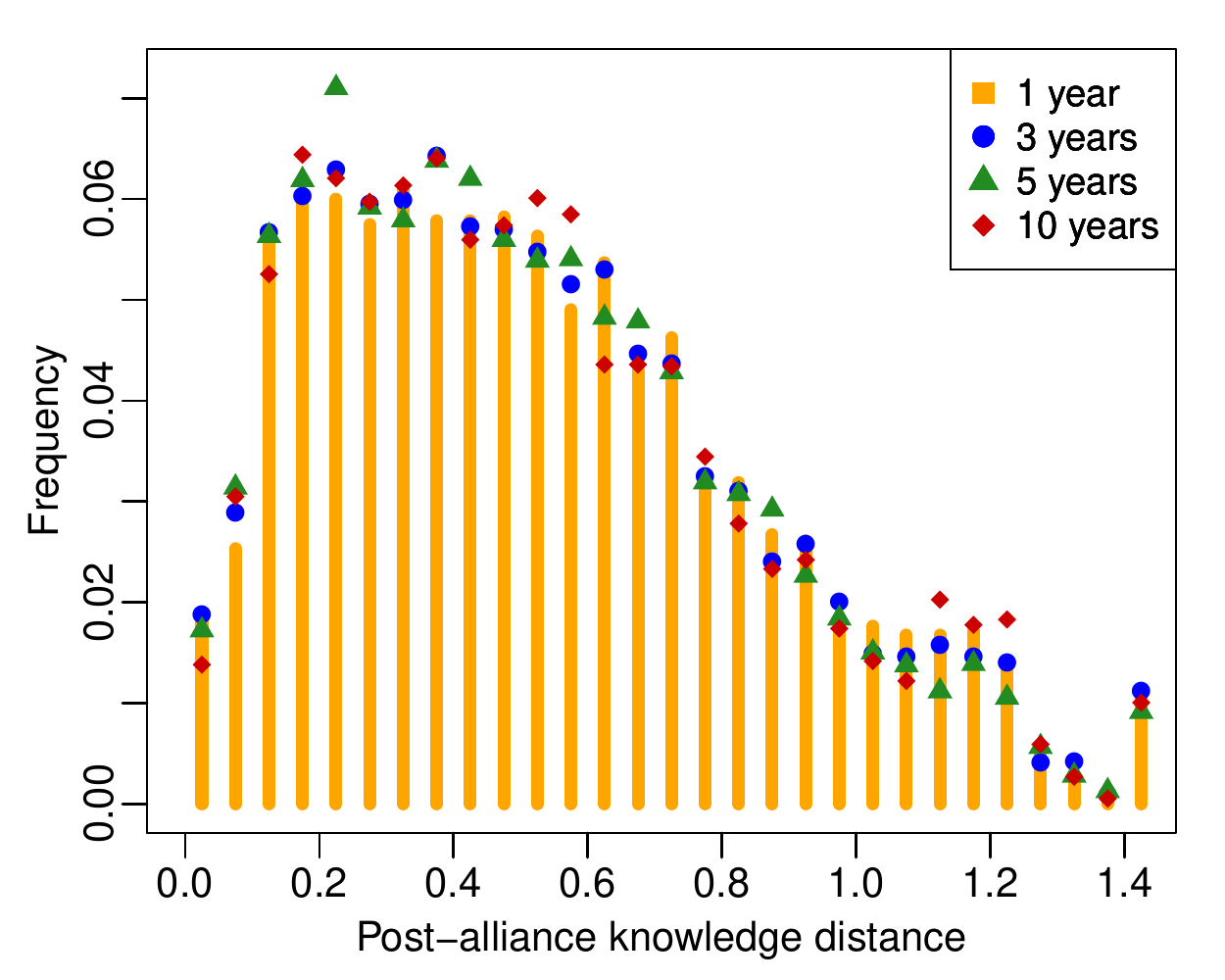}
(b)
\includegraphics[width=0.45\textwidth,angle=0]{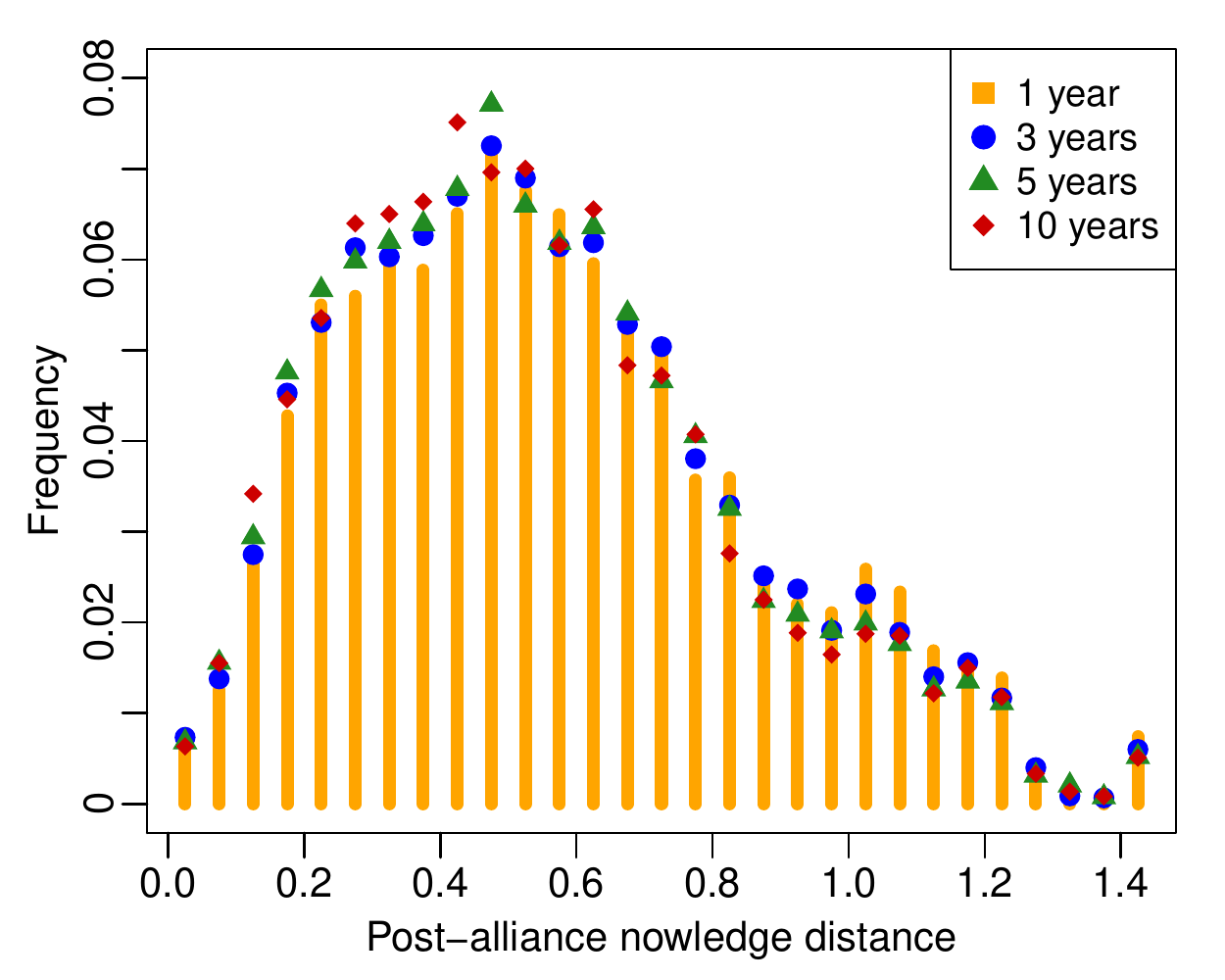}
\end{center}
\caption{Empirical knowledge distance between every pair of partnered firms, computed 1, 3, 5 and 10 years after the date of the alliance formation. 
In (a) we have calculated the distance using the 8 dimensional knowledge space defined by the IPC scheme and in (b) used the 35 dimensional knowledge space defined by the ISI-OST-INPI classification scheme.}
\label{fig:patent_empirical_postalliance_distr}
\end{figure}
In the first step, we analyze the \emph{empirical distribution of knowledge distances} for different alliance durations. 
The NBER patent data set has a time-granularity of 1 year. 
This forces us to use time increments of 1 year with a minimum window of 1 year. 
In Fig. \ref{fig:patent_empirical_postalliance_distr} we show the post-alliance knowledge distance distribution for different time windows: 1, 3, 5 and 10 years.
We find that, for both knowledge space metrics, the shape of the knowledge distance distribution appears to have the same shape, irrespective of the time window chosen. 
This allows for two conclusions. 
First, an assumed increase of the alliance duration does not considerably impact the post-alliance distance distribution, most likely because firms do not move much in knowledge space over time. 
Second, because of this our modeling approach is robust against the (unknown) duration of alliances.
There is a firm relation between $\tau$ and $\mu$ as discussed in Figure \ref{fig:comprehensive_model_goodness}. 
But even for larger durations $\tau$, the properly calibrated model can be used to reproduce the empirical distribution of post-alliance knowledge distances. 


\paragraph{Changes of knowledge distances}

\begin{figure}[h!]
\begin{center}
(a)
\includegraphics[width=0.45\textwidth,angle=0]{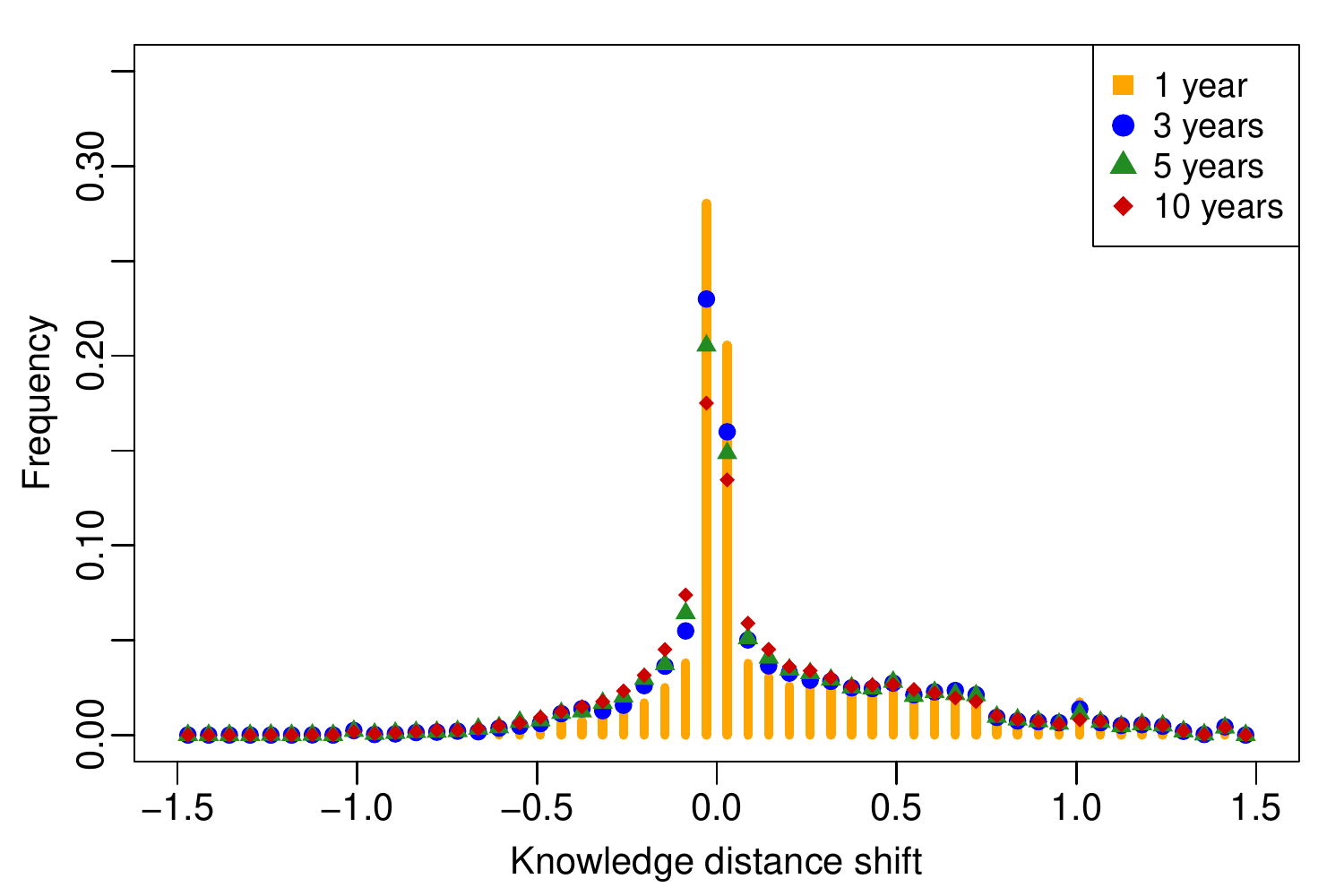}
(b)
\includegraphics[width=0.45\textwidth,angle=0]{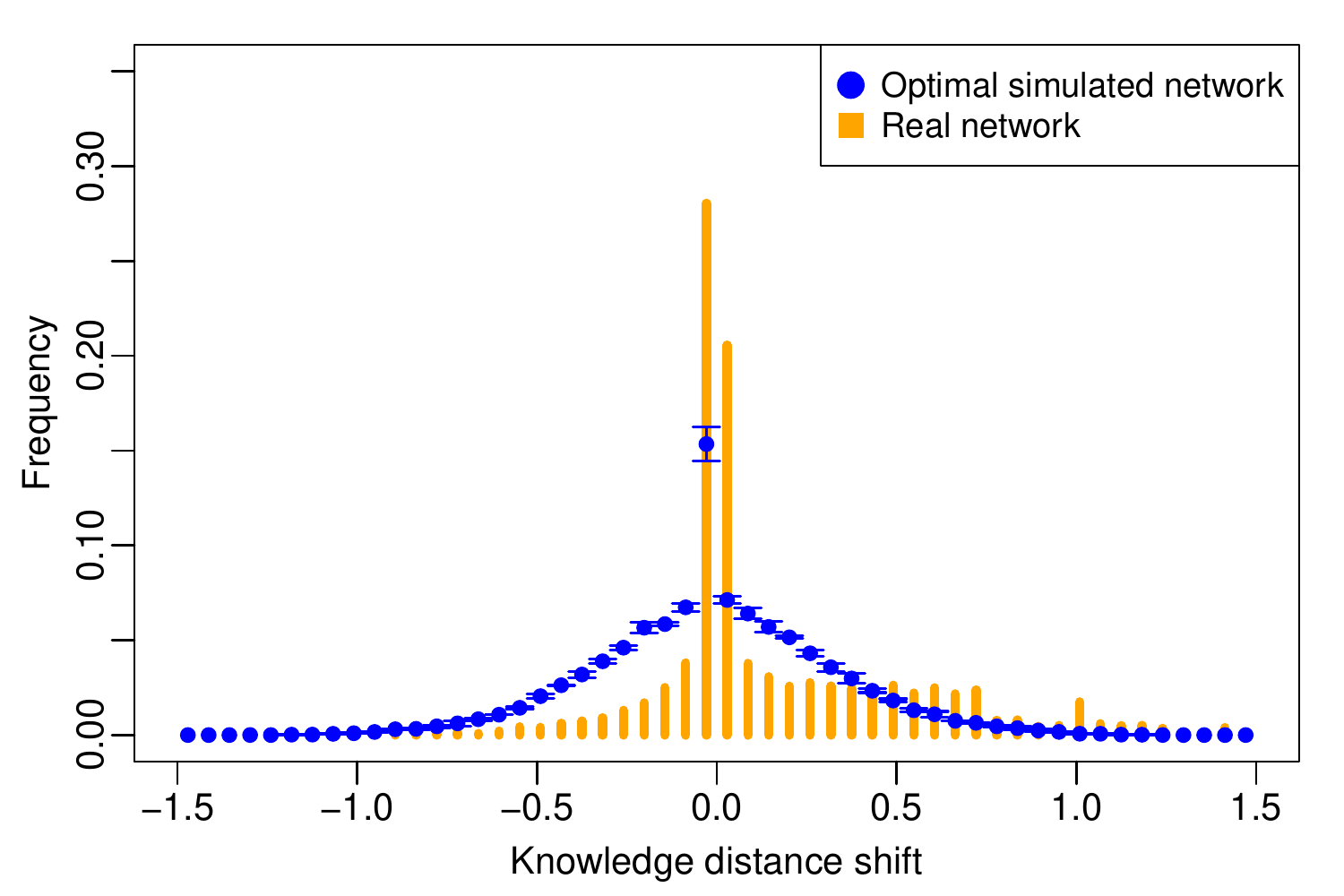}
\end{center}
\caption[Empirical shifts of knowledge distance.]{(a) Empirical shift of knowledge distance between every pair of partnered firms, computed 1, 3, 5 and 10 years after the date of the alliance formation.
(b) Empirical and simulated distance shifts between all allied firms for $\tau=700$days and $\mu=10^{-7}$days$^{-1}$ . In both plots, we report results obtained considering the 35 dimensional space defined by the ISI-OST-INPI classification scheme.}
\label{fig:patent_empirical_dist_shift}
\end{figure}
In the second step, we calculate the \emph{changes of the knowledge distances} between the empirical \emph{pre-alliance} distance distribution shown in Figure \ref{fig:patent_empirical_prealliance_distr} and the empirical \emph{post-alliance} distance distribution shown in Figure \ref{fig:patent_empirical_postalliance_distr}. 
Because the time of alliance termination is not known, we have to vary the duration again in time steps of 1 year. 
Our results are shown in Fig. \ref{fig:patent_empirical_dist_shift} (a) for the ISI-OST-INPI classification scheme. 
The results for the IPC scheme are rather similar and therefore not shown here. 

There are two remarkable observations in Fig. \ref{fig:patent_empirical_dist_shift} (a). 
First, the distributions are clearly centered around zero, i.e. small changes of knowledge distances are very frequent. 
Larger changes of knowledge distances are rare, but not unlikely. 
This is in line with the broad distributions we find for both the pre- and the post-alliance knowledge distances. 
Second, the results for the changes in knowledge distances are robust against choosing a longer duration for alliances. 
We note that positive changes are more prominently seen for the ISI-OST-INPI classification scheme, whereas they look symmetric for the IPC scheme.

In order to see whether these findings are captured by our model of knowledge exchange, we have calculated the changes in distances also in the computer simulations (using optimal parameters). 
The result is shown in Fig. \ref{fig:patent_empirical_dist_shift}(b), where we compare the changes in the  empirical knowledge distances (also shown on the left side) with the changes in the simulated knowledge distances. 
We see that the (rather) symmetric distribution peaked at zero can be reproduced by our model, even with the long tails. 
Some deviations occur close to zero, where the empirical distribution is more peaked, to decay faster than the simulated one. 
These deviations are in line with the deviations already discussed for Fig. \ref{fig:patent_empirical_postalliance_distr}, where small distances are slightly overrepresented in the simulated initial conditions. 

More interesting is the fact that both the empirical and the simulated distributions of distance changes exhibit tails on \emph{both} sides.
I.e., some alliances cause the partners to significantly move \textit{closer} in the knowledge space, whilst during other alliances the partners significantly move \textit{farther away}. 
In our model of knowledge exchange, however, we have only assumed that alliance partners \emph{approach} each other in knowledge space, which would lead to a left skew distribution of (mostly negative) changes.
The explanation comes from the fact that firms, while forming new alliances, can be still engaged in existing alliances or establish new ones. 
The resulting change in the knowledge distance with respect to a given partner is thus the superposition of all influences a firm is subject to, at the time of alliance termination. 
In other words, there exists a nonlinear (and nontrivial) feedback of the network formation process on the knowledge exchange dynamics, which we further comment on in Sect. \ref{sec:conclusions}. 
At this point, we just emphasize that this influence is correctly captured in our agent-based model, as it also reflects the movement of agents farther away in knowledge space.

\section{Estimating the performance of knowledge exchange}
\label{sec:performance}


One of the most prominent reasons for R\&D collaborations, seen from the perspective of the firm, is the exchange of knowledge, as already argued in Section \ref{sec:intro}. 
The formation of R\&D alliances between individual firms results in a large-scale R\&D network pictured in Fig. \ref{fig:network}. 
This network represents one projection of the systemic, or ``macroscopic'', perspective.
The complementary projection of the systemic perspective is given by the  knowledge space made up by the patent portfolios of individual firms. 
Only the dimensions of that space are defined by the (two different) patent classification schemes. 
Firms collectively shape, and explore, this knowledge space by forming alliances and exchanging knowledge with their partners. 

The collective exploration of the knowledge space is beneficial for the whole system \citep{dosi02:_exploitation_exploration_innovation}. 
Therefore, we now want to evaluate the performance of this collective exploration, by analyzing different indicators. 
We do not intend to directly match these indicators to any possible empirical counterpart. 
Rather, we address the question of to what extent the empirical R\&D network corresponds to a simulated network that is optimal with respect to such indicators. 

As the possibly simplest performance indicator for our simulated networks, we investigate the \emph{total distance} that all agents have traveled in knowledge space \citep{tomasello2016knowledge_exchange}. 
For an individual agent, the length $L_{i}(t)$ of the path traveled in the knowledge space  is defined by the sum of all distances that the agent traveled in every time step of the simulation until time $t$:
\begin{equation} 
\label{eq:knowledge_path}
L_{i}(T_{\mathrm{max}}) = \int_{t=0}^{T_{\mathrm{max}}} \abs{\dot{\mathbf{x}_{i}}(t)} \, \mathrm{d}t
\end{equation}
For our convenience $T_{\mathrm{max}}$ is the duration of the entire computer simulation. 
It should be noted that the measure $\abs{\dot{\mathbf{x}_{i}}(t)}\, \mathrm{d}t$ is a positive scalar and expresses the actual \emph{distance} traveled by the agent $i$, differently from its \emph{net displacement} $\dot{\mathbf{x}_{i}}(t)\, \mathrm{d}t$, which is a vectorial quantity. 

\begin{figure}[htbp]
(a)  \includegraphics[width=0.29\textwidth]{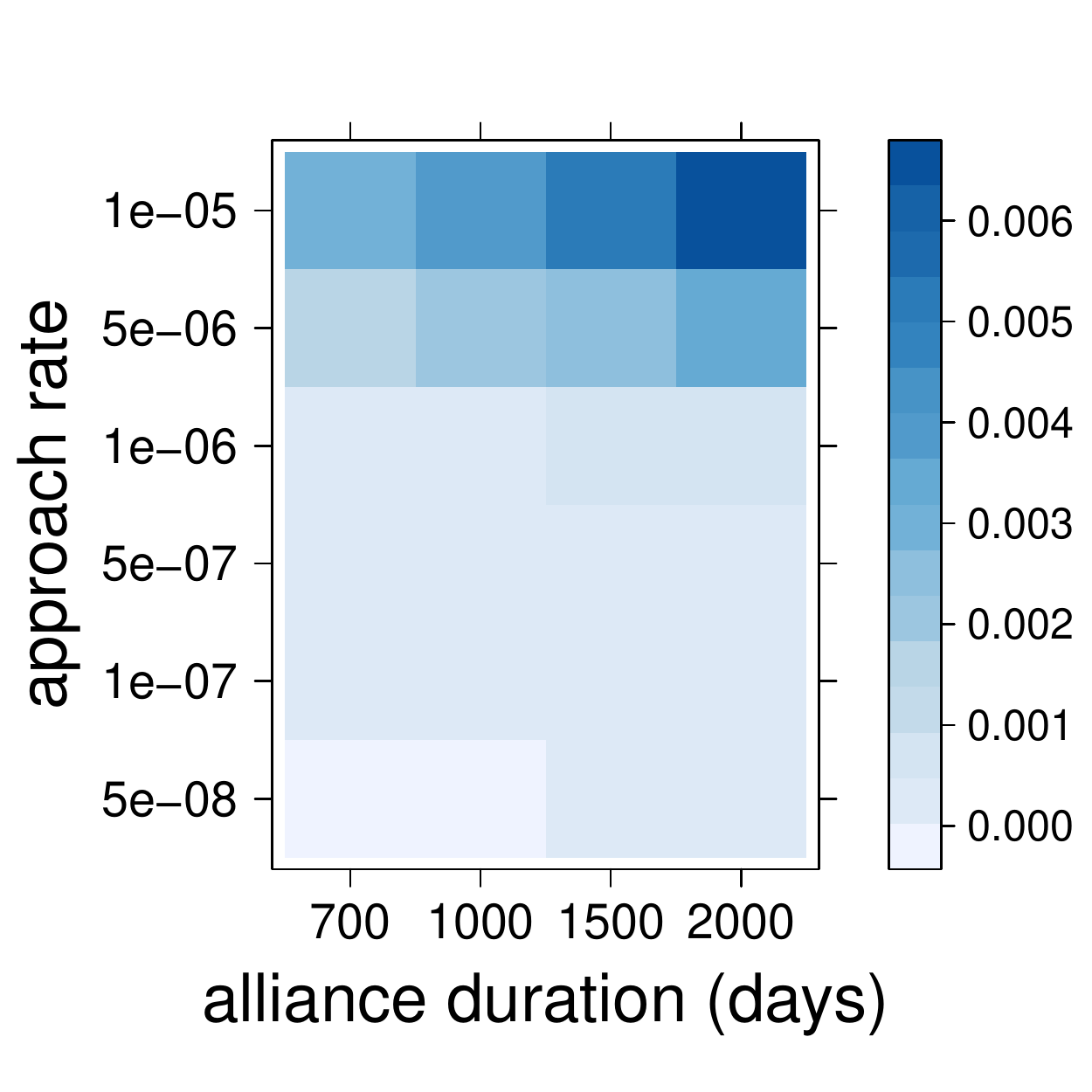}
\hfill
(b)  \includegraphics[width=0.29\textwidth]{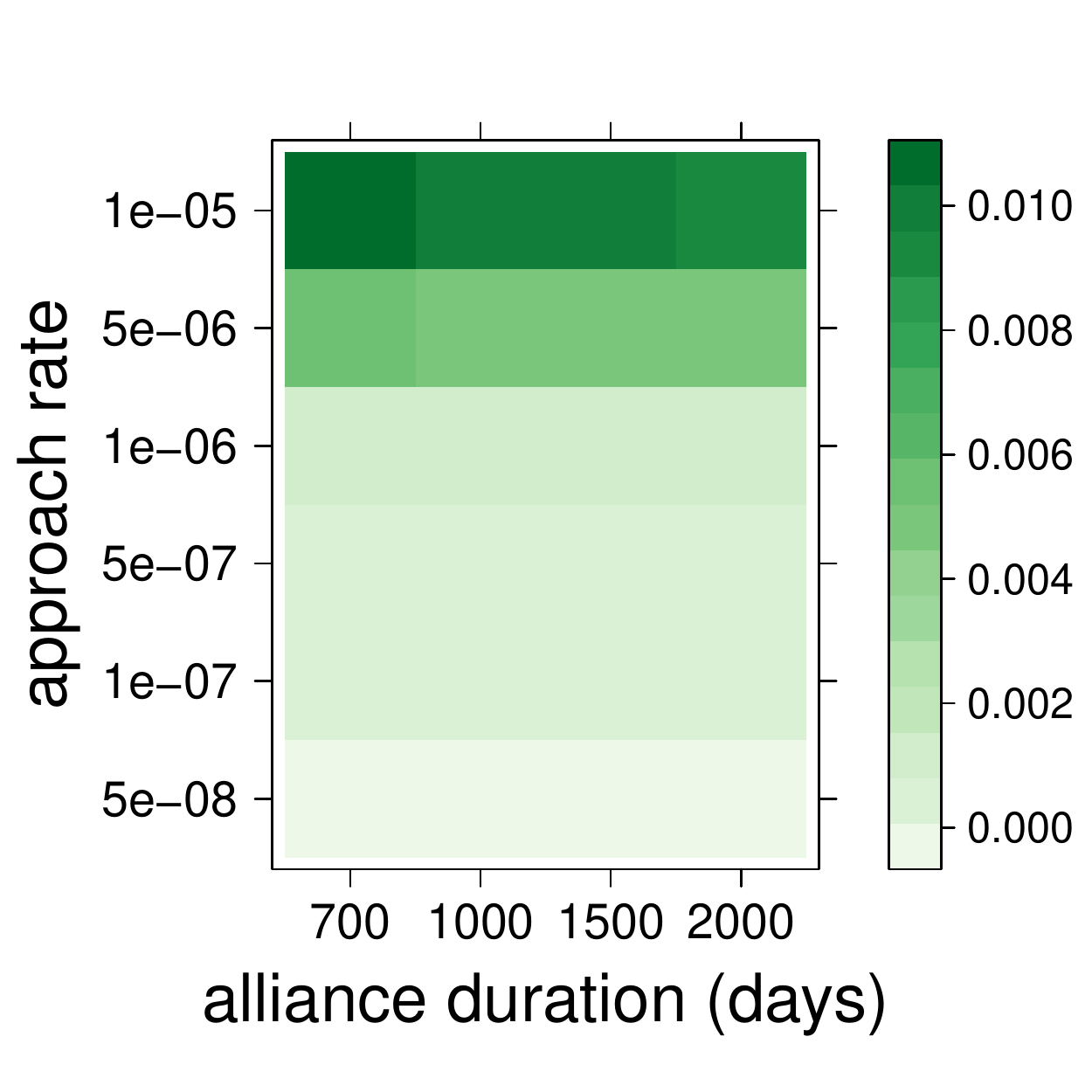}
\hfill
(c) \includegraphics[width=0.29\textwidth]{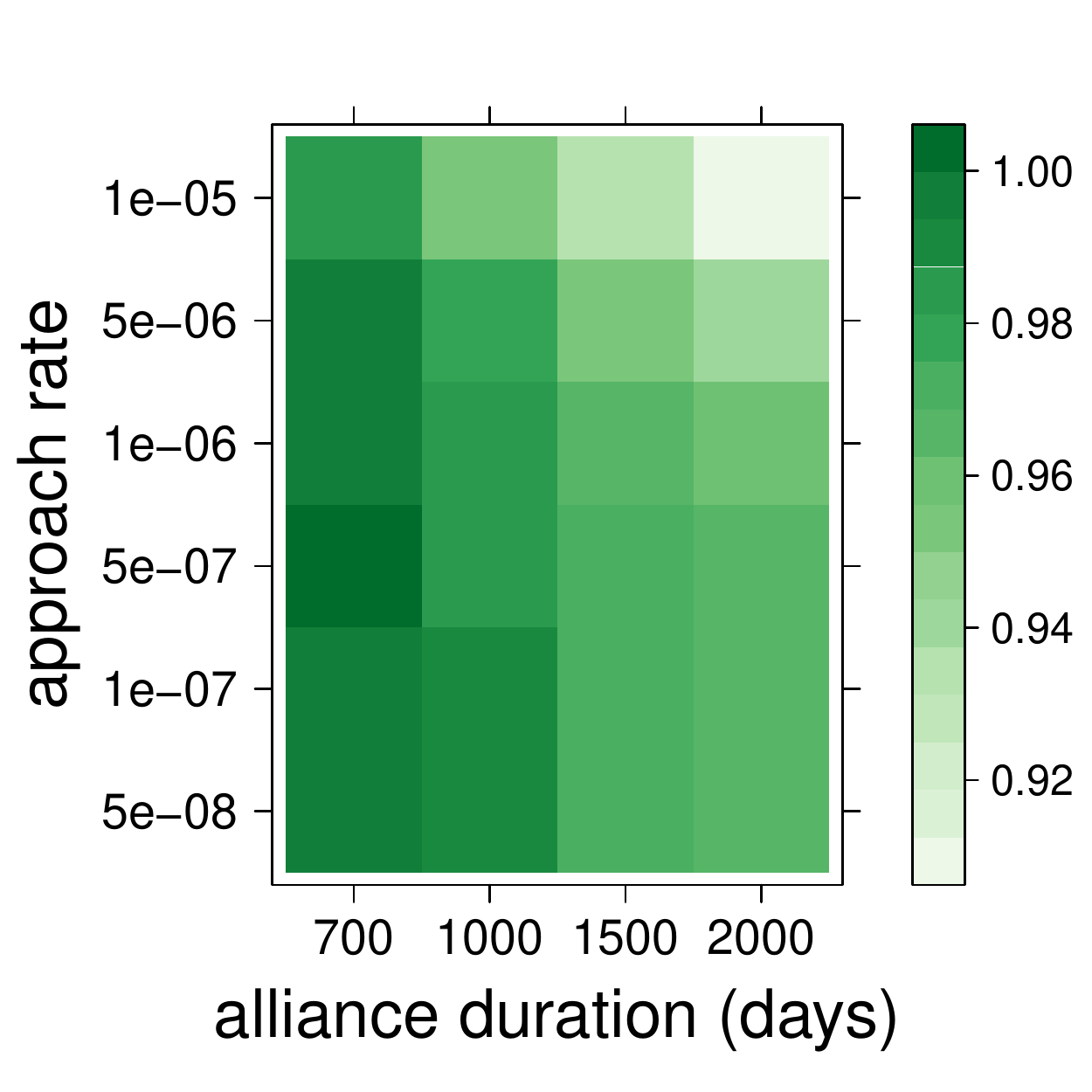}
  \caption{The heatmap for the average total distance, $\mean{L}$, traveled by the agents is reported in (a). In (b) we report the heatmap for network collaboration efficiency, $\mathcal{C}$, and in (c) the heatmap for its normalized and rescaled version version, $\mathcal{\hat{C}}_{n}$. For all the three plots, we report results obtained using the 35 dimensional space defined by the ISI-OST-INPI classification scheme.}
  \label{fig:knowledge_path}
\end{figure}
The measure $L_{i}(t)$ is then averaged over all the agents in the network to obtain the averaged total distance in knowledge space, $\mean{L(t)}=N^{-1} \sum_i^N{L_i(t)}$.
This is shown in Fig. \ref{fig:knowledge_path}(a) as a heat map of the bi-dimensional $(\tau,\mu)-$parameter space.  
We argue that  a higher value of $\mean{L}$, i.e. a better exploration of the knowledge space, corresponds to a higher systemic performance, because, as already discussed in Section \ref{sec:intro_knowledge_exchange}, firms are proven to innovate more when they come in contact with more technological opportunities. 

At the same time, using $\mean{L}$ as performance indicator does not give us detailed insights because, as 
Fig. \ref{fig:knowledge_path}(a) shows, higher approach rates $\mu$ always lead to larger distances traveled in knowledge space, for any alliance duration $\tau$. 
This motivates us to propose a more refined performance indicator, $\mathcal{C}$, that also takes into account 
the number of active collaborations, $k_{i}^{\mathrm{act}}(t)$, that cause firms to move in knowledge space at a given time $t$. 
I.e. in our model $k_i^{\mathrm{act}}(t)$ is the \emph{degree} of agent $i$ at time $t$. 
We remind that not all collaborations are active at a given time; some are terminated and become inactive after a characteristic time $\tau$. 
As firms engaged in alliances incur in costs, we consider that $\mathcal{C}$ should decrease with increasing number of active collaborations:
\begin{equation}
  \label{eq:collaboration_performance}
  {\cal{C}} = \; \int_{t=0}^{T_{\mathrm{max}}} {\frac{\sum_{i=1}^N{\abs{\dot{\mathbf{x}_{i}}(t)}}} { \sum_{i=1}^N{k_i^{\mathrm{act}}(t)}} \:\mathrm{d}t}
  \; = \; \frac{1}{2}
  \int_{t=0}^{T_{\mathrm{max}}} {\frac{\sum_{i=1}^N{\abs{\dot{\mathbf{x}_{i}}(t)}}}
{M^{\mathrm{act}}(t)} \:\mathrm{d}t }
\end{equation}
$\cal{C}$ is called \textit{collaboration efficiency} because it considers how much output, i.e.  movement in knowledge space, the system achieves for a given input, covering e.g. the costs to maintain collaboration links.  
The ratio of the two sums in Eq. \ref{eq:collaboration_performance} gives the total distance traveled \emph{per active collaboration} during a given time step $dt$. 
This ratio is then integrated over the duration $T_{\mathrm{max}}$ of the simulation, to obtain the overall collaboration performance $\cal{C}$ of the network. 
The sum of all agents' degrees, $\sum_i k_i^{\mathrm{act}}(t) = 2\cdot M^{\mathrm{act}}(t)$, gives us twice the total number of active links, $M^{\mathrm{act}}(t)$, in the network at time $t$ because every link connects two agents. 
By plugging this into Eq. \ref{eq:collaboration_performance}, we obtain the second expression for the collaboration efficiency. 
It means that, given equal total knowledge distances $ \sum_i^N{L_i(t)}$, an R\&D network with less alliances would explore the knowledge space more efficiently. 

We use Eq. \ref{eq:collaboration_performance} to compute the collaboration efficiency $\cal{C}$ for every network generated during the simulations.
$\cal{C}$ is then averaged over 100 simulations for every combination of parameters. 
The results are shown in the heat map of Fig. \ref{fig:knowledge_path}(b) for simulations using the 35 dimensional knowledge space, where we plot the collaboration efficiency $\mathcal{C}$ against the two parameters characterizing the knowledge exchange, exchange rate $\mu$ and alliance duration $\tau$. 
Comparing this to Fig. \ref{fig:knowledge_path}(a), we find again that $\mu$ has a strong impact, i.e. the larger the knowledge exchange rate, the better the performance. 
However, the influence of $\tau$ has reversed. 
Now, performance increases with shorter alliance durations, which is understandable because we take the costs of alliances into account. 
The larger $\tau$, the more alliances exist concurrently and have to be maintained. 
This causes the overall performance to drop.

\begin{figure}[h]
\begin{center}
\includegraphics[width=
0.5\textwidth,angle=0]{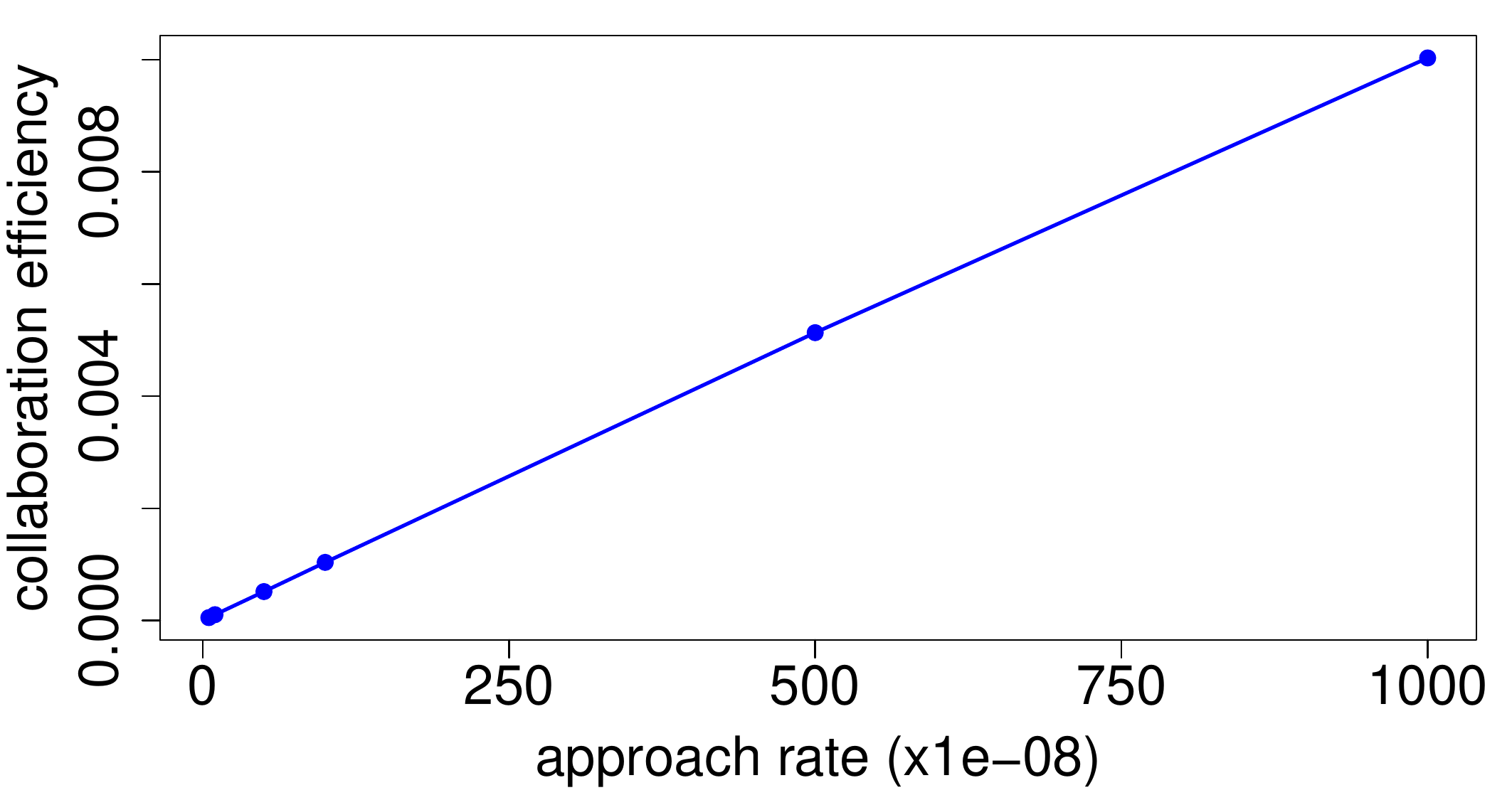}
\end{center}
\caption{Collaboration efficiency $\mathcal{C}$ dependent on the knowledge exchange rate $\mu$ for a fixed alliance duration of $\tau=700$ days. The knowledge of the agents was embedded in the 35 dimensional space defined by the ISI-OST-INPI classification scheme.}
\label{fig:comprehensive_model_mean_knowledge_path}
\end{figure}
To further investigate the strong impact of $\mu$, we plot in  Fig. \ref{fig:comprehensive_model_mean_knowledge_path}  for a fixed alliance duration $\tau=$700 days how the collaboration efficiency $\mathcal{C}$ changes with the knowledge exchange rate.
We find that there is a linear relation between these two quantities (similar for other values of $\tau$, not shown). 
This is agreement with the definition of $\mathcal{C}$, Eq. \eqref{eq:collaboration_performance}, where the leading term of the numerator is linear in $\mu$. 
Non-linear terms of the order $O(\mu^2)$ play a less important role since $\mu$ is small.
Hence, for a better comparison of the collaboration efficiency across different values of $\tau$, we rescale $\mathcal{C}$ as $\mathcal{\hat C}=\mathcal{C}/\mu$.
In addition, to obtain a dimensionless quantity that varies between 0 and 1, we normalize $\mathcal{\hat{C}}$ by its maximum value obtained for a given set of parameters $\mu$, $\tau$, i.e.
\begin{equation}
\label{eq:2}
\mathcal{\hat C}_{n}=\frac{\cal{\hat{C}}}{\max_{\mu,\tau}\cal{\hat{C}}}=
\frac{\cal{C/\mu}}{\max_{\mu,\tau}({\cal{C}}/{\mu})}
\end{equation}
In Fig.\ref{fig:knowledge_path}(c), we show $\mathcal{\hat C}_{n}$ for all combinations of the knowledge exchange parameters.  
We confirm, even after normalization, the tendency that the performance increases with smaller values of $\tau$,  i.e., for the range of parameters considered the best value of $\tau$ is 700 days. 
But for the knowledge exchange rate, we obtain a more detailed and heterogeneous dependency.
Given $\tau=$700 days, the optimal value of $\mu$ is now at $5\times 10^{-7}$ days$^{-1}$.





In conclusion, we find that the highest efficiency in knowledge exchange is obtained for medium exchange rates and short alliance durations. 
These results are found by means of computer simulations of our model. 
In order to transfer such insights to firms in real R\&D networks, some restrictions apply. 

It is understandable that a shorter collaboration is more beneficial because it implies, as already mentioned, that in a given time interval a smaller number of concurrent alliances exist. 
A reduced number of collaborations, on the other hand, allows a firm to move efficiently along one or a few directions in the knowledge space. 

In order to keep the performance of exploring the knowledge space high, firms have to compensate 
shorter alliance durations  by  larger knowledge exchange rates. 
While this is feasible in our model, it may not hold under practical circumstances because
firms have limits of how much new knowledge they can absorb at a given time. 
So, there are upper limits for the knowledge exchange rate $\mu$.  

On the other hand, it is obvious that there is a lower bound for an optimal alliance duration $\tau$. 
Firms have to get to know each other, and have to establish procedures of collaborations which takes time. 
Hence, organizational and management arguments suggest that $\tau$ cannot simply approach zero, also because the knowledge exchange rate cannot simply be increased to arbitrary large values.

Such arguments apply when choosing \emph{realistic ranges} of parameters $\tau$ and $\mu$ in our model. 
Thus, via the choice of parameters our model takes these limitations into account. 
In addition, it is useful to understand the \emph{impact} of these model parameters on the systemic performance in knowledge exploration. 
As we have shown, there is a nonlinear, and non-trivial, relation between knowledge exchange rate $\mu$ and alliance duration $\tau$. 
With an increasing alliance duration, more links become active at the same time, thus forcing firms to cope with the effect of multiple partnerships. 
This results in a reduced motion, i.e. a reduced collective exploration, in the knowledge space.
In other words, the density of the collaboration network increases with $\tau$ and, after a certain threshold, the addition of a new link has a negative marginal effect on the overall exploration of the knowledge space. 


\section{Discussion and conclusions}
\label{sec:conclusions}

This paper aims at a \textit{quantitative} understanding of knowledge exchange in R\&D networks. 
``Quantitative'' means, 
(i) we propose a model that reflects the two tightly connected processes of forming R\&D alliances and knowledge exchange, 
(ii) we analyze large-scale data sets capturing R\&D alliances and knowledge bases of firms to calibrate the model parameters, 
(iii) we perform extensive computer simulations to analyze the performance of knowledge exchange in R\&D network. 
Instead of repeating our findings, in this section we highlight a few points for further discussion. 

\paragraph{Partner selection and network formation}

We have proposed an agent-based model that consists of two interlinked phases: (1) the formation of the R\&D network, which is called the \emph{exploration phase} because agents explore the social capital of potential partners, and (2) the exchange of knowledge on the formed network, which is called the \emph{exploitation phase} because agents utilize the collaboration with partners to move in knowledge space. 

The calibration of our model against real data was performed through a two-step procedure. 
By means of an alliance data set, we have estimated a set of link probabilities that allow us to reproduce the topology of the R\&D collaboration network. 
Subsequently, through a second data set on firm patents, we have estimated parameters for the knowledge exchange between firms  and the duration of R\&D alliances.

For the formation of the R\&D network, we found that firms exhibit a strong tendency to connect to network incumbents. 
Precisely, 65\% of the collaborations initiated by incumbents, as well as a surprising 90\% of the collaborations initiated by newcomers, are addressed to another incumbent.
In this regard, the validation of our model brings additional support to the theory of the importance of existing network structures in the formation of new R\&D collaborations \citep[see][]{podolny1993status, raub1990reputation}.


\paragraph{Dynamics of knowledge exchange}

Because the model part related to the network formation was already investigated by \citet{tomasello2014therole}, in this paper we mainly focus on modeling knowledge exchange as a motion of agents in a predefined knowledge space. 
The knowledge base of agents is estimated by the patent portfolio of firms. 
Therefore, the dimensionality of the knowledge space is given by the patent classifications for which we use two different schemes, (a) IPC and (b) ISI-OST-INPI. With respect to our model, their difference is mainly in the \emph{number} of dimensions, (a) 8 and (b) 35.
Thus, we can also address the question how an expansion of the number of dimensions of the knowledge space affects the results of our model. 

Firms are characterized by a position in this knowledge space, which changes over time as they obtain new patents. 
As the focus of our paper is on R\&D \emph{collaborations}, the model does not assume that firms can change their position by independent R\&D activities. 
But we have indirectly covered this  by the fact that, in our model, each time a new alliance starts agents get assigned a new position if they are not already involved in existing alliances. 
Differently from the model introduced by \citet{tomasello2016knowledge_exchange}, where the motion of every agent was driven by only one partner at every time step, in the present model the agents are subject to a motion resulting from interactions with multiple partners. 
As we have already discussed in Sect.~\ref{sec:comprehensive_model_knowledge}, our dynamics assumes that knowledge exchange causes agents to \emph{approach} each other in knowledge space, not just in one dimension but in all dimensions. 
This takes into account the effect of knowledge spillovers that go beyond the exchange of very specific knowledge.  

Analyzing \emph{empirically} the impact of R\&D collaborations on firms' knowledge positions, we found that small changes in knowledge distances are dominating the dynamics in knowledge space (see Fig. \ref{fig:patent_empirical_dist_shift}).
I.e., real firms do not significantly change their knowledge positions as a consequence of their collaborations.  
This supports our conclusion that \emph{most} alliances exert only a \emph{weak} influence on the knowledge positions of firms.
However, we also find that
\emph{some} (non-negligible) alliances are able to cause a \emph{strong} movement in knowledge space.

\paragraph{Interplay between network formation and knowledge exchange}

It is an interesting observation that the empirical distribution of distance changes is rather \emph{symmetric} with respect to zero; although we note that positive changes are more prominently seen in the ISI-OST-INPI classification scheme  (see Fig. \ref{fig:patent_empirical_dist_shift}).
This means that, in the period elapsed during a specific R\&D alliance, firms not only approach each other in knowledge space (negative distance changes) but also move farther away (positive distance changes). 

This finding can be also reproduced by our agent-based model, which is remarkable because there we assume only that agents approach each other. 
However, the model of knowledge exchange considers the \emph{combined impact} of all interactions on the knowledge position of an agent. 
Our model can reproduce both negative and positive distance changes because they result not only from the knowledge dynamics, but also from the \emph{network dynamics}. 
This means that, while being engaged in one alliance, agents start to form new alliances with other partners which can drive them away from their current partners. 
Hence, it is the complex interplay between network formation and knowledge exchange that at the end can explain the collective exploration of the knowledge space. 


\paragraph{Pre- and post-alliance distance distributions}

For the calibration of our knowledge exchange dynamics, special attention was given to the knowledge distances between firms at two points in time, at the moment of alliance formation (which is known) and at the moment of alliance termination (which is not known).
Hence, the alliance duration $\tau$ is considered as one free parameter of our model. 

We emphasize that in our model proximity in knowledge space is \emph{not} a precondition for agents to form alliances. 
Consequently, distances can be quite \emph{large}, which is in line with the empirical fact that the distribution of \emph{pre-alliance distances} is clearly left-skewed (see Fig. \ref{fig:patent_empirical_prealliance_distr}). 
On the other hand, we have also shown that the most \textit{frequent} pre-alliance distance between firms are shorter than the one expected at random (see Fig. \ref{fig:emp-sim-pre}). 
The most probable value (i.e. the maximum of the distribution) is clearly different from zero and could be interpreted as an optimal distance in knowledge space for firms to engage in an alliance.

In our model, we have taken the distribution of pre-alliance distances as an input, i.e. we have sampled the knowledge positions of agents that are \emph{not} engaged in an alliance at that time from this distribution. 
Agents that \emph{are} in an alliance at that time, however, keep the knowledge position simulated by the model. 
The combined procedure of sampling knowledge positions has two advantages: first, we retain information about the similarity  of collaborating firms in the knowledge space. 
For example, if firms from the same industrial sector were more likely to have an alliance, this would be captured in the pre-alliance distance distribution (e.g. smaller alliance distances are more probable) and considered in our model.
Second, by using the empirical knowledge vectors, we also keep information about the technological areas in which firms usually file patents.
Thus, we partially account for the size of firm portfolios of patents.




Regarding  the distribution of  \emph{post-alliance distances}, we have shown that it is not really different from the distribution of pre-alliance distances, which reflects the fact that most changes in knowledge positions are rather small. 
This finding holds for both patent classification schemes, i.e. it is robust against the number of dimensions of the knowledge space. 
It is also robust against the assumed alliance duration (see Fig. \ref{fig:patent_empirical_postalliance_distr}). 

So, if firms do not move much in knowledge space, why is their position important? 
Firms rather use the available information about knowledge positions of their partners to establish new collaborations. 
Therefore, a firm's position in knowledge space is more a \emph{determinant} than a consequence of its R\&D alliances.

In our model, we have used the distribution of  post-alliance distances to compare the outcome of our  simulations with their empirical counterpart. 
Using optimized parameters for the simulated network formation, we vary the parameters for knowledge exchange to find the best match between the empirical and the simulation post-alliance distance distribution (see Fig. \ref{fig:comprehensive_model_goodness}). 
As the result, we obtain the values $\mu=1\times 10^{-7}, .., 5\times 10^{-7}$ for the knowledge exchange rate and $\tau=700$ for the alliance duration. 
$\mu$ has a relatively low value, which is in line with the fact that most firms do not move much in knowledge space, while $\tau$ indicates a characteristic duration of around two years (700 days). 
The latter finding is consistent with our previous theoretical assumptions and a number of previous studies \citep[see][]{inkpen01:_why_do_some_strat_allian, phelps2003technological}.
We note that these optimal parameters for knowledge exchange are obtained from a procedure that compares \emph{simulation} and \emph{empirics}. 
In the following, we discuss that we have derived the same optimal parameters from a pure simulation approach, using assumptions about performance. 

\paragraph{Performance of knowledge exchange}

In this paper, we are not only interested in the \emph{dynamics} of knowledge exchange in R\&D networks, but also in the \emph{performance}.
The latter we define as a systemic property, i.e. we do not discuss the performance of individual firms, but the collective performance of the whole R\&D network in efficiently exploring the knowledge space. 

The dynamics assumed for knowledge exchange would suggest that higher knowledge exchange rates $\mu$ and longer alliance durations $\tau$ are always better for exploration. 
This, however, implies that firms cope with many concurrent alliances at the same time and have an infinite capacity of absorbing new knowledge.
A more realistic scenario has to take into account that alliances are also costly, i.e. establishing and maintaining concurrent alliances is constrained by capacities. 
To capture these influences, we have proposed the (normalized) \emph{collaboration efficiency} $\mathcal{\hat{C}}_{n}$, Eqs. \eqref{eq:collaboration_performance}, \eqref{eq:2}, as a new performance measure. 
Analyzing how $\mathcal{\hat{C}}_{n}$ depends on the parameters for knowledge exchange $\mu$ and $\tau$, we find that the collaboration efficiency is maximized for values $\mu=5\times 10^{-7}$ and $\tau=700$ (see Fig. \ref{fig:knowledge_path}c), which match the above given optimized parameters from Fig. \ref{fig:comprehensive_model_goodness}. 
Because this result was found by comparing only simulations, we regard this as an independent way to confirm the parameters found by comparing the empirical and the simulated distribution of post-alliance distances.  
This means that, using our approach, it is possible to obtain a configuration that is both \textit{realistic} and \textit{optimized} with respect to the collaboration performance. 

When discussing these findings, we already pointed out that in real-world applications the parameters $\mu$ and $\tau$ are rather determined by the firm's abilities to quickly establish a collaboration and to absorb new knowledge fast. 
Hence, organizational and managerial constraints apply, which should be considered for choosing values for these parameters.  

Nevertheless, with our model we are able to point toward policies aimed at system optimization.
Effective policies to obtain an improved collaboration network would incentivize {short} R\&D alliances and higher knowledge exchange rates.
Practically, it would be impossible to directly enforce shorter alliance durations or faster learning rates.
But measures could include, for instance, rewards for co-patenting activities if  these are carried out as early as possible after the establishment of an R\&D alliance.
The goal would be to stimulate companies to explore other knowledge positions with new partners while limiting the duration of a single alliance and to avoid having too many active collaborations at the same time.

\bigskip

In conclusion, we argue that our model can successfully reproduce both network-related and knowledge-related features of a real inter-organizational R\&D network.
At the same time, our data-driven approach provides a unique method to estimate the systemic performance of R\&D collaborations. 
We note that our model is extendable to other collaboration systems, beyond the domain of R\&D networks, provided that the agents can be unequivocally positioned in a knowledge space. 
Our approach thus contributes to a comprehensive understanding of the effects of knowledge exchange in  dynamically evolving collaboration networks.

\section*{Compliance with Ethical Standards}

GV acknowledges support  from  the  Swiss  State  Secretariat for  Education,  Research  and  Innovation  (SERI),  Grant No.  C14.0036 as well as from EU COST Action TD1210 KNOWeSCAPE.
M.~V.~T. acknowledges financial support from the Seed Project SP-RC 01-15 ``Performance and resilience of collaboration networks'', granted by the ETH Zurich Risk Center. 
CJT acknowledges financial support from the University Research Priority Program on Social Network, University of Zurich.
The authors declare that they have no conflict of interest.

\bibliographystyle{sg-bibstyle}
\bibliography{all_references_MT.bib}

\end{document}